\pdfoutput=1
\documentclass[11pt,twoside,a4paper,pdftex,cmspaper,final,collab]{cms-tdr}

\begin{document}\cmsNoteHeader{CFT-09-009}
%
%
%

%
%
\hyphenation{env-iron-men-tal}
\hyphenation{had-ron-i-za-tion}
\hyphenation{cal-or-i-me-ter}
\hyphenation{de-vices}
%
%
\RCS$Revision: 1.62 $
\RCS$Date: 2010/01/01 21:27:22 $
\RCS$Name:  $
%
%
%

\newcommand {\etal}{\mbox{et al.}\xspace} 
\newcommand {\ie}{\mbox{i.e.}\xspace}     
\newcommand {\eg}{\mbox{e.g.}\xspace}     
\newcommand {\etc}{\mbox{etc.}\xspace}     
\newcommand {\vs}{\mbox{\sl vs.}\xspace}      
\newcommand {\mdash}{\ensuremath{\mathrm{-}}} 

\newcommand {\Lone}{Level-1\xspace} 
\newcommand {\Ltwo}{Level-2\xspace}
\newcommand {\Lthree}{Level-3\xspace}

\providecommand{\ACERMC} {\textsc{AcerMC}\xspace}
\providecommand{\ALPGEN} {{\textsc{alpgen}}\xspace}
\providecommand{\CHARYBDIS} {{\textsc{charybdis}}\xspace}
\providecommand{\CMKIN} {\textsc{cmkin}\xspace}
\providecommand{\CMSIM} {{\textsc{cmsim}}\xspace}
\providecommand{\CMSSW} {{\textsc{cmssw}}\xspace}
\providecommand{\COBRA} {{\textsc{cobra}}\xspace}
\providecommand{\COCOA} {{\textsc{cocoa}}\xspace}
\providecommand{\COMPHEP} {\textsc{CompHEP}\xspace}
\providecommand{\EVTGEN} {{\textsc{evtgen}}\xspace}
\providecommand{\FAMOS} {{\textsc{famos}}\xspace}
\providecommand{\GARCON} {\textsc{garcon}\xspace}
\providecommand{\GARFIELD} {{\textsc{garfield}}\xspace}
\providecommand{\GEANE} {{\textsc{geane}}\xspace}
\providecommand{\GEANTfour} {{\textsc{geant4}}\xspace}
\providecommand{\GEANTthree} {{\textsc{geant3}}\xspace}
\providecommand{\GEANT} {{\textsc{geant}}\xspace}
\providecommand{\HDECAY} {\textsc{hdecay}\xspace}
\providecommand{\HERWIG} {{\textsc{herwig}}\xspace}
\providecommand{\HIGLU} {{\textsc{higlu}}\xspace}
\providecommand{\HIJING} {{\textsc{hijing}}\xspace}
\providecommand{\IGUANA} {\textsc{iguana}\xspace}
\providecommand{\ISAJET} {{\textsc{isajet}}\xspace}
\providecommand{\ISAPYTHIA} {{\textsc{isapythia}}\xspace}
\providecommand{\ISASUGRA} {{\textsc{isasugra}}\xspace}
\providecommand{\ISASUSY} {{\textsc{isasusy}}\xspace}
\providecommand{\ISAWIG} {{\textsc{isawig}}\xspace}
\providecommand{\MADGRAPH} {\textsc{MadGraph}\xspace}
\providecommand{\MCATNLO} {\textsc{mc@nlo}\xspace}
\providecommand{\MCFM} {\textsc{mcfm}\xspace}
\providecommand{\MILLEPEDE} {{\textsc{millepede}}\xspace}
\providecommand{\ORCA} {{\textsc{orca}}\xspace}
\providecommand{\OSCAR} {{\textsc{oscar}}\xspace}
\providecommand{\PHOTOS} {\textsc{photos}\xspace}
\providecommand{\PROSPINO} {\textsc{prospino}\xspace}
\providecommand{\PYTHIA} {{\textsc{pythia}}\xspace}
\providecommand{\SHERPA} {{\textsc{sherpa}}\xspace}
\providecommand{\TAUOLA} {\textsc{tauola}\xspace}
\providecommand{\TOPREX} {\textsc{TopReX}\xspace}
\providecommand{\XDAQ} {{\textsc{xdaq}}\xspace}

\newcommand {\DZERO}{D\O\xspace}     


\newcommand{\de}{\ensuremath{^\circ}}
\newcommand{\ten}[1]{\ensuremath{\times \text{10}^\text{#1}}}
\newcommand{\unit}[1]{\ensuremath{\text{\,#1}}\xspace}
\newcommand{\mum}{\ensuremath{\,\mu\text{m}}\xspace}
\newcommand{\micron}{\ensuremath{\,\mu\text{m}}\xspace}
\newcommand{\cm}{\ensuremath{\,\text{cm}}\xspace}
\newcommand{\mm}{\ensuremath{\,\text{mm}}\xspace}
\newcommand{\mus}{\ensuremath{\,\mu\text{s}}\xspace}
\newcommand{\keV}{\ensuremath{\,\text{ke\hspace{-.08em}V}}\xspace}
\newcommand{\MeV}{\ensuremath{\,\text{Me\hspace{-.08em}V}}\xspace}
\newcommand{\GeV}{\ensuremath{\,\text{Ge\hspace{-.08em}V}}\xspace}
\newcommand{\TeV}{\ensuremath{\,\text{Te\hspace{-.08em}V}}\xspace}
\newcommand{\PeV}{\ensuremath{\,\text{Pe\hspace{-.08em}V}}\xspace}
\newcommand{\keVc}{\ensuremath{{\,\text{ke\hspace{-.08em}V\hspace{-0.16em}/\hspace{-0.08em}c}}}\xspace}
\newcommand{\MeVc}{\ensuremath{{\,\text{Me\hspace{-.08em}V\hspace{-0.16em}/\hspace{-0.08em}c}}}\xspace}
\newcommand{\GeVc}{\ensuremath{{\,\text{Ge\hspace{-.08em}V\hspace{-0.16em}/\hspace{-0.08em}c}}}\xspace}
\newcommand{\TeVc}{\ensuremath{{\,\text{Te\hspace{-.08em}V\hspace{-0.16em}/\hspace{-0.08em}c}}}\xspace}
\newcommand{\keVcc}{\ensuremath{{\,\text{ke\hspace{-.08em}V\hspace{-0.16em}/\hspace{-0.08em}c}^\text{2}}}\xspace}
\newcommand{\MeVcc}{\ensuremath{{\,\text{Me\hspace{-.08em}V\hspace{-0.16em}/\hspace{-0.08em}c}^\text{2}}}\xspace}
\newcommand{\GeVcc}{\ensuremath{{\,\text{Ge\hspace{-.08em}V\hspace{-0.16em}/\hspace{-0.08em}c}^\text{2}}}\xspace}
\newcommand{\TeVcc}{\ensuremath{{\,\text{Te\hspace{-.08em}V\hspace{-0.16em}/\hspace{-0.08em}c}^\text{2}}}\xspace}

\newcommand{\pbinv} {\mbox{\ensuremath{\,\text{pb}^\text{$-$1}}}\xspace}
\newcommand{\fbinv} {\mbox{\ensuremath{\,\text{fb}^\text{$-$1}}}\xspace}
\newcommand{\nbinv} {\mbox{\ensuremath{\,\text{nb}^\text{$-$1}}}\xspace}
\newcommand{\percms}{\ensuremath{\,\text{cm}^\text{$-$2}\,\text{s}^\text{$-$1}}\xspace}
\newcommand{\lumi}{\ensuremath{\mathcal{L}}\xspace}
\newcommand{\Lumi}{\ensuremath{\mathcal{L}}\xspace}
%
\newcommand{\LvLow}  {\ensuremath{\mathcal{L}=\text{10}^\text{32}\,\text{cm}^\text{$-$2}\,\text{s}^\text{$-$1}}\xspace}
\newcommand{\LLow}   {\ensuremath{\mathcal{L}=\text{10}^\text{33}\,\text{cm}^\text{$-$2}\,\text{s}^\text{$-$1}}\xspace}
\newcommand{\lowlumi}{\ensuremath{\mathcal{L}=\text{2}\times \text{10}^\text{33}\,\text{cm}^\text{$-$2}\,\text{s}^\text{$-$1}}\xspace}
\newcommand{\LMed}   {\ensuremath{\mathcal{L}=\text{2}\times \text{10}^\text{33}\,\text{cm}^\text{$-$2}\,\text{s}^\text{$-$1}}\xspace}
\newcommand{\LHigh}  {\ensuremath{\mathcal{L}=\text{10}^\text{34}\,\text{cm}^\text{$-$2}\,\text{s}^\text{$-$1}}\xspace}
\newcommand{\hilumi} {\ensuremath{\mathcal{L}=\text{10}^\text{34}\,\text{cm}^\text{$-$2}\,\text{s}^\text{$-$1}}\xspace}


\newcommand{\zp}{\ensuremath{\mathrm{Z}^\prime}\xspace}


\newcommand{\kt}{\ensuremath{k_{\mathrm{T}}}\xspace}
\newcommand{\BC}{\ensuremath{{B_{\mathrm{c}}}}\xspace}
\newcommand{\bbarc}{\ensuremath{{\overline{b}c}}\xspace}
\newcommand{\bbbar}{\ensuremath{{b\overline{b}}}\xspace}
\newcommand{\ccbar}{\ensuremath{{c\overline{c}}}\xspace}
\newcommand{\JPsi}{\ensuremath{{J}/\psi}\xspace}
\newcommand{\bspsiphi}{\ensuremath{B_s \to \JPsi\, \phi}\xspace}
\newcommand{\AFB}{\ensuremath{A_\mathrm{FB}}\xspace}
\newcommand{\EE}{\ensuremath{e^+e^-}\xspace}
\newcommand{\MM}{\ensuremath{\mu^+\mu^-}\xspace}
\newcommand{\TT}{\ensuremath{\tau^+\tau^-}\xspace}
\newcommand{\wangle}{\ensuremath{\sin^{2}\theta_{\mathrm{eff}}^\mathrm{lept}(M^2_\mathrm{Z})}\xspace}
\newcommand{\ttbar}{\ensuremath{{t\overline{t}}}\xspace}
\newcommand{\stat}{\ensuremath{\,\text{(stat.)}}\xspace}
\newcommand{\syst}{\ensuremath{\,\text{(syst.)}}\xspace}

\newcommand{\HGG}{\ensuremath{\mathrm{H}\to\gamma\gamma}}
\newcommand{\gev}{\GeV}
\newcommand{\GAMJET}{\ensuremath{\gamma + \mathrm{jet}}}
\newcommand{\PPTOJETS}{\ensuremath{\mathrm{pp}\to\mathrm{jets}}}
\newcommand{\PPTOGG}{\ensuremath{\mathrm{pp}\to\gamma\gamma}}
\newcommand{\PPTOGAMJET}{\ensuremath{\mathrm{pp}\to\gamma +
\mathrm{jet}
}}
\newcommand{\MH}{\ensuremath{\mathrm{M_{\mathrm{H}}}}}
\newcommand{\RNINE}{\ensuremath{\mathrm{R}_\mathrm{9}}}
\newcommand{\DR}{\ensuremath{\Delta\mathrm{R}}}


\newcommand{\PT}{\ensuremath{p_{\mathrm{T}}}\xspace}
\newcommand{\pt}{\ensuremath{p_{\mathrm{T}}}\xspace}
\newcommand{\ET}{\ensuremath{E_{\mathrm{T}}}\xspace}
\newcommand{\HT}{\ensuremath{H_{\mathrm{T}}}\xspace}
\newcommand{\et}{\ensuremath{E_{\mathrm{T}}}\xspace}
\newcommand{\Em}{\ensuremath{E\!\!\!/}\xspace}
\newcommand{\Pm}{\ensuremath{p\!\!\!/}\xspace}
\newcommand{\PTm}{\ensuremath{{p\!\!\!/}_{\mathrm{T}}}\xspace}
\newcommand{\ETm}{\ensuremath{E_{\mathrm{T}}^{\mathrm{miss}}}\xspace}
\newcommand{\MET}{\ensuremath{E_{\mathrm{T}}^{\mathrm{miss}}}\xspace}
\newcommand{\ETmiss}{\ensuremath{E_{\mathrm{T}}^{\mathrm{miss}}}\xspace}
\newcommand{\VEtmiss}{\ensuremath{{\vec E}_{\mathrm{T}}^{\mathrm{miss}}}\xspace}

%

\newcommand{\ga}{\ensuremath{\gtrsim}}
\newcommand{\la}{\ensuremath{\lesssim}}
\newcommand{\swsq}{\ensuremath{\sin^2\theta_W}\xspace}
\newcommand{\cwsq}{\ensuremath{\cos^2\theta_W}\xspace}
\newcommand{\tanb}{\ensuremath{\tan\beta}\xspace}
\newcommand{\tanbsq}{\ensuremath{\tan^{2}\beta}\xspace}
\newcommand{\sidb}{\ensuremath{\sin 2\beta}\xspace}
\newcommand{\alpS}{\ensuremath{\alpha_S}\xspace}
\newcommand{\alpt}{\ensuremath{\tilde{\alpha}}\xspace}

\newcommand{\QL}{\ensuremath{Q_L}\xspace}
\newcommand{\sQ}{\ensuremath{\tilde{Q}}\xspace}
\newcommand{\sQL}{\ensuremath{\tilde{Q}_L}\xspace}
\newcommand{\ULC}{\ensuremath{U_L^C}\xspace}
\newcommand{\sUC}{\ensuremath{\tilde{U}^C}\xspace}
\newcommand{\sULC}{\ensuremath{\tilde{U}_L^C}\xspace}
\newcommand{\DLC}{\ensuremath{D_L^C}\xspace}
\newcommand{\sDC}{\ensuremath{\tilde{D}^C}\xspace}
\newcommand{\sDLC}{\ensuremath{\tilde{D}_L^C}\xspace}
\newcommand{\LL}{\ensuremath{L_L}\xspace}
\newcommand{\sL}{\ensuremath{\tilde{L}}\xspace}
\newcommand{\sLL}{\ensuremath{\tilde{L}_L}\xspace}
\newcommand{\ELC}{\ensuremath{E_L^C}\xspace}
\newcommand{\sEC}{\ensuremath{\tilde{E}^C}\xspace}
\newcommand{\sELC}{\ensuremath{\tilde{E}_L^C}\xspace}
\newcommand{\sEL}{\ensuremath{\tilde{E}_L}\xspace}
\newcommand{\sER}{\ensuremath{\tilde{E}_R}\xspace}
\newcommand{\sFer}{\ensuremath{\tilde{f}}\xspace}
\newcommand{\sQua}{\ensuremath{\tilde{q}}\xspace}
\newcommand{\sUp}{\ensuremath{\tilde{u}}\xspace}
\newcommand{\suL}{\ensuremath{\tilde{u}_L}\xspace}
\newcommand{\suR}{\ensuremath{\tilde{u}_R}\xspace}
\newcommand{\sDw}{\ensuremath{\tilde{d}}\xspace}
\newcommand{\sdL}{\ensuremath{\tilde{d}_L}\xspace}
\newcommand{\sdR}{\ensuremath{\tilde{d}_R}\xspace}
\newcommand{\sTop}{\ensuremath{\tilde{t}}\xspace}
\newcommand{\stL}{\ensuremath{\tilde{t}_L}\xspace}
\newcommand{\stR}{\ensuremath{\tilde{t}_R}\xspace}
\newcommand{\stone}{\ensuremath{\tilde{t}_1}\xspace}
\newcommand{\sttwo}{\ensuremath{\tilde{t}_2}\xspace}
\newcommand{\sBot}{\ensuremath{\tilde{b}}\xspace}
\newcommand{\sbL}{\ensuremath{\tilde{b}_L}\xspace}
\newcommand{\sbR}{\ensuremath{\tilde{b}_R}\xspace}
\newcommand{\sbone}{\ensuremath{\tilde{b}_1}\xspace}
\newcommand{\sbtwo}{\ensuremath{\tilde{b}_2}\xspace}
\newcommand{\sLep}{\ensuremath{\tilde{l}}\xspace}
\newcommand{\sLepC}{\ensuremath{\tilde{l}^C}\xspace}
\newcommand{\sEl}{\ensuremath{\tilde{e}}\xspace}
\newcommand{\sElC}{\ensuremath{\tilde{e}^C}\xspace}
\newcommand{\seL}{\ensuremath{\tilde{e}_L}\xspace}
\newcommand{\seR}{\ensuremath{\tilde{e}_R}\xspace}
\newcommand{\snL}{\ensuremath{\tilde{\nu}_L}\xspace}
\newcommand{\sMu}{\ensuremath{\tilde{\mu}}\xspace}
\newcommand{\sNu}{\ensuremath{\tilde{\nu}}\xspace}
\newcommand{\sTau}{\ensuremath{\tilde{\tau}}\xspace}
\newcommand{\Glu}{\ensuremath{g}\xspace}
\newcommand{\sGlu}{\ensuremath{\tilde{g}}\xspace}
\newcommand{\Wpm}{\ensuremath{W^{\pm}}\xspace}
\newcommand{\sWpm}{\ensuremath{\tilde{W}^{\pm}}\xspace}
\newcommand{\Wz}{\ensuremath{W^{0}}\xspace}
\newcommand{\sWz}{\ensuremath{\tilde{W}^{0}}\xspace}
\newcommand{\sWino}{\ensuremath{\tilde{W}}\xspace}
\newcommand{\Bz}{\ensuremath{B^{0}}\xspace}
\newcommand{\sBz}{\ensuremath{\tilde{B}^{0}}\xspace}
\newcommand{\sBino}{\ensuremath{\tilde{B}}\xspace}
\newcommand{\Zz}{\ensuremath{Z^{0}}\xspace}
\newcommand{\sZino}{\ensuremath{\tilde{Z}^{0}}\xspace}
\newcommand{\sGam}{\ensuremath{\tilde{\gamma}}\xspace}
\newcommand{\chiz}{\ensuremath{\tilde{\chi}^{0}}\xspace}
\newcommand{\chip}{\ensuremath{\tilde{\chi}^{+}}\xspace}
\newcommand{\chim}{\ensuremath{\tilde{\chi}^{-}}\xspace}
\newcommand{\chipm}{\ensuremath{\tilde{\chi}^{\pm}}\xspace}
\newcommand{\Hone}{\ensuremath{H_{d}}\xspace}
\newcommand{\sHone}{\ensuremath{\tilde{H}_{d}}\xspace}
\newcommand{\Htwo}{\ensuremath{H_{u}}\xspace}
\newcommand{\sHtwo}{\ensuremath{\tilde{H}_{u}}\xspace}
\newcommand{\sHig}{\ensuremath{\tilde{H}}\xspace}
\newcommand{\sHa}{\ensuremath{\tilde{H}_{a}}\xspace}
\newcommand{\sHb}{\ensuremath{\tilde{H}_{b}}\xspace}
\newcommand{\sHpm}{\ensuremath{\tilde{H}^{\pm}}\xspace}
\newcommand{\hz}{\ensuremath{h^{0}}\xspace}
\newcommand{\Hz}{\ensuremath{H^{0}}\xspace}
\newcommand{\Az}{\ensuremath{A^{0}}\xspace}
\newcommand{\Hpm}{\ensuremath{H^{\pm}}\xspace}
\newcommand{\sGra}{\ensuremath{\tilde{G}}\xspace}
\newcommand{\mtil}{\ensuremath{\tilde{m}}\xspace}
\newcommand{\rpv}{\ensuremath{\rlap{\kern.2em/}R}\xspace}
\newcommand{\LLE}{\ensuremath{LL\bar{E}}\xspace}
\newcommand{\LQD}{\ensuremath{LQ\bar{D}}\xspace}
\newcommand{\UDD}{\ensuremath{\overline{UDD}}\xspace}
\newcommand{\Lam}{\ensuremath{\lambda}\xspace}
\newcommand{\Lamp}{\ensuremath{\lambda'}\xspace}
\newcommand{\Lampp}{\ensuremath{\lambda''}\xspace}
\newcommand{\spinbd}[2]{\ensuremath{\bar{#1}_{\dot{#2}}}\xspace}

\newcommand{\MD}{\ensuremath{{M_\mathrm{D}}}\xspace}
\newcommand{\Mpl}{\ensuremath{{M_\mathrm{Pl}}}\xspace}
\newcommand{\Rinv} {\ensuremath{{R}^{-1}}\xspace}

%
%
\hyphenation{en-viron-men-tal}

\cmsNoteHeader{CFT-09-009}
\title{Performance of the CMS Hadron Calorimeter with  Cosmic Ray Muons and LHC Beam Data
}

\date{\today}

\author{The CMS Collaboration}

\abstract{
The CMS Hadron Calorimeter in the barrel, endcap and forward regions is fully commissioned.
Cosmic ray data were taken with and without magnetic field at the surface hall and after installation in the
experimental hall, hundred meters underground.
Various measurements were also performed during the few days of beam in the LHC in September 2008.
Calibration parameters were extracted, and the energy response of the HCAL determined from test beam
data has been checked.
}

\hypersetup{
pdfauthor={Mayda M. Velasco, Vasken Hagopian},%
pdftitle={Performance of the CMS Hadron Calorimeter with  Cosmic Ray Muons and LHC Beam Data},%
pdfsubject={CMS},%
pdfkeywords={CMS, detector, calorimeter, HCAL, performance}}

\maketitle 

\section{Introduction}

The primary goal of the Compact Muon Solenoid\,(CMS) experiment~\cite{bib:det}
is to explore particle physics at the TeV energy scale exploiting the
proton-proton
collisions delivered by the Large Hadron Collider\,(LHC)~\cite{bib:lhc}.
This paper describes the
performance of the CMS  Hadron Calorimeter\,(HCAL).  The measurement of the
detector response started with the characterization and calibration of
representative samples of all detector components
in the laboratory using a test beam~\cite{bib:1,bib:2,bib:3,bib:4}.
After assembly of the detector, the calibration was improved and
validated using cosmic ray muons and LHC beam. The cosmic ray muon data also allowed the testing of the whole data acquisition process,  up to the event
reconstruction and analysis. The validation includes comparisons with Monte Carlo simulation studies,
which are fundamental in understanding the detailed performance of the detector.

CMS has taken cosmic ray muon data with and without magnetic field at the
surface hall in 2006, during the Magnet Test and Cosmic Challenge\,(MTCC)
described in Ref.~\cite{bib:mtcc}. Recently, data were taken in the underground
cavern at $\sim 100$\,m depth with the magnet off in 2007\,(CRUZET: Cosmic RUn at ZEro Tesla)
and with the magnet on in 2008\,(CRAFT: Cosmic Run At Four Tesla\,\cite{bib:craft}).

In September 2008, the LHC delivered a single circulating beam of 450\,\GeV/$c$ protons for a few days.
In addition, as  part of the commissioning of the LHC, the beam was also sent onto
collimator targets creating  accelerator produced muons.
These ``beam splash'' events proved very
useful to determine the relative timing of various the HCAL systems~\cite{bib:8}, as well as to  validate
the performance of several components.

Knowledge of the HCAL performance  acquired during CRAFT and with beam splash data
will prove crucial to the understanding of  data in the early stages of LHC operation.
This will allow the timely analysis of benchmark QCD processes
and of known missing energy signals. Any search for new phenomena must be grounded on
these benchmark analyses.

The paper is structured as follows. Section~\ref{sec:det}
provides a  brief description of the Hadron Calorimeter. Section~\ref{sec:magnet} discusses the impact
of the magnetic field on the various parts of the HCAL subsystems.   The
characteristics  of the basic behavior of all the HCAL subsystems and the optimization
of key hardware settings are summarized in Section~\ref{sec:char}.
Finally, Section~\ref{sec:validation} discusses
the calibration improvements obtained with the various data samples.

\section{CMS Hadron Calorimeter}
\label{sec:det}
The main CMS subsystems, in increasing radial distance from the beam intersection point\,(IP), are
the silicon pixel and  silicon strip tracker, the electromagnetic calorimeter, the  hadron
calorimeter, and finally the muon system.
In addition, in the forward direction there are two specialized calorimeters~\cite{bib:det}
that did not participate in the cosmic ray  runs and are not discussed in this paper.
The detector is symmetric for positive and negative $z$ coordinates, the $z$ axis being along the
beam direction with $z=0$ at the IP.
The polar angle, $\theta$, is  measured from the $z$ direction,
and the azimuthal angle, $\phi$, is  measured in the transverse plane.
The pseudorapidity is $\eta \equiv - \ln \tan (\theta/2)$.

\begin{figure}[hbtp]
  \begin{center}
    \includegraphics[width=0.8\textwidth]{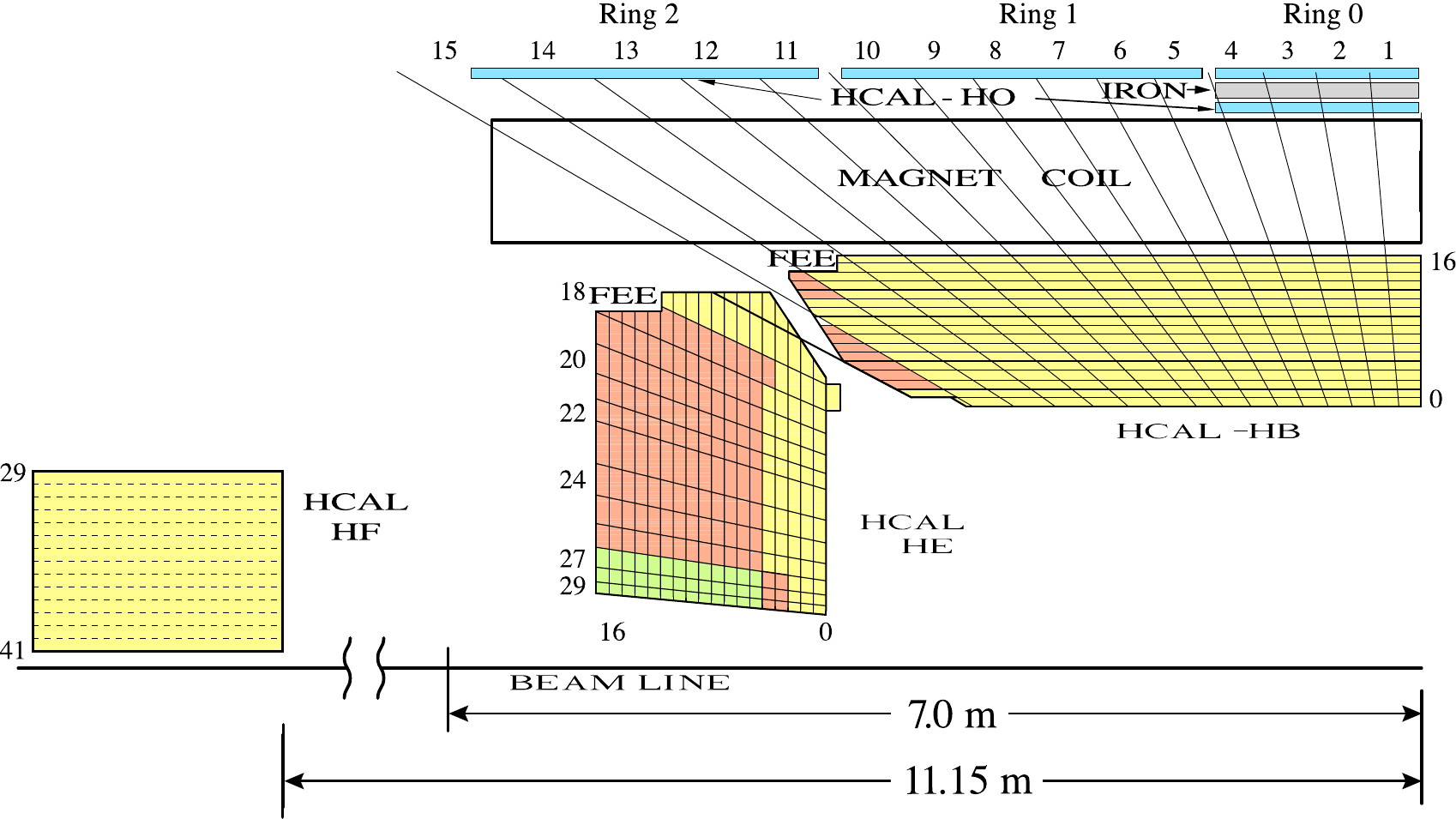}
    \hspace{1cm}
    \caption{The CMS HCAL detector\,(quarter slice).
``FEE''  indicates
the locations of the Front End Electronics for HB and HE.
The signals of the tower segments
with the same color are added optically, to provide the HCAL ``longitudinal'' segmentation.
HB, HE and HF are built of 36 identical azimuthal
wedges\,($\Delta \phi$ = 20 degrees).
}
    \label{fig:hcal_layout}
  \end{center}
\end{figure}

The HCAL includes four distinct subsystems: the barrel\,(HB), endcap\,(HE),
outer\,(HO) and forward\,(HF) calorimeters (see Fig.~\ref{fig:hcal_layout}). HB and HE,  which  are
inside the cryostat of the CMS superconducting solenoid that provides a 3.8\,T magnetic field,
are sampling calorimeters where the absorber material is brass and the active material is
scintillator. The sampling
fraction is about 7\%. HB and HE are separated by a gap which is located approximately
at a polar angle of 57 degrees but is not projective to the center of CMS, in order
 to minimize the effect of the uninstrumented gap.
HB covers the $|\eta|$ range  from zero  to approximately 1.4; the
$|\eta|$ range between 1.3 and 1.4 is  shared by HB and HE; HE covers $|\eta|$
from 1.3 to 3.0.  HB is built of 18  wedges, each of
 which covers 20 degrees in $\phi$, and are divided in
5 degree sectors.
HE is made of brass disks, interleaved with scintillator wedges which cover 20 degrees in $\phi$,
which in turn are divided in four 5 degree sectors.
Because of the space constraint within the magnet
cryostat, the HB thickness is limited to 5.8\,hadronic interaction
lengths at $\eta=0$ and increases to 10 interaction lengths at
$|\eta|=1.2$.
To catch the energy leakage from   HB, layers of scintillators are
placed outside the solenoid cryostat: they  constitute  HO.
About 5\% of all hadrons above 100\,GeV deposit energy in HO.
In $\phi$, HO has a 12-fold structure, with each 30 degree component being divided in six 5
degree sectors.  In $\eta$, HO is composed of five ``rings'',
 which follow the structure of the magnet
return yoke and of the muon chambers.
Ring~0 covers the $\eta$ range between -0.35 and 0.35,
 Rings~$\pm$1 cover the $|\eta|$
 range between 0.35 and 0.87, and Rings~$\pm$2 cover the $|\eta|$ range between
0.87 and 1.2.
The  quartz fiber and steel HF calorimeter, with fibers parallel to the beam direction, covers the forward region of $|\eta|$, between 3.0 and 5.2.
HF is constructed in wedges of 20
degrees and each wedge contains two $\phi$ sectors of 10 degrees.
The calorimeter tower segmentation in $\eta$ and $\phi$ of HB, HE and HO subsystems
is  0.087$\times$0.087 except in HE
for $|\eta|$ above 1.74, where the $\eta$ segmentation ranges from
0.09 to 0.35 and the $\phi$ segmentation is 0.175.
The HF segmentation is 0.175$\times$0.175 except for $|\eta|$ above 4.7,
 where the segmentation is 0.175$\times$0.35.

Figure~\ref{fig:hcal_layout} shows a  schematic quarter view of the hadron calorimeter system in the barrel,
endcap and forward regions. Also shown are the locations of some of the Front End Electronics\,(FEE). The HF FEEs
(not shown) are placed  around a ring at $|\eta|=3$\,(tower number 29) and HO FEEs are located inside the muon detectors at
various locations. Each HB and HE tower  has 17 scintillator layers, except near the overlap region between HB and HE.
Each scintillator tile of a tower is read out by an embedded wavelength shifting fiber and the signals
are added optically. The color scheme in Fig.~\ref{fig:hcal_layout} denotes the
longitudinal segmentation of the read out; all layers shown with the same color in one $\eta$ tower are summed.
The optical signals for HB, HE and HO are detected by hybrid photodiodes\,(HPD) with 19 independent pixels; 18 for read out of
fibers and one for monitoring. The
HPDs are designed to work inside the magnetic field, provided their axes are aligned with the magnetic field.
This alignment is adequate for HB and HE, but it was found that the
 HO HPDs were off by as much as 40 degrees.
This misalignment was due to the difficulty in simulating the magnetic field inside the return yoke.
Therefore, an effort is under way to study the possible replacement of the HO HPDs with silicon
photomultipliers that are insensitive to magnetic fields and
have a better signal to noise discrimination.

In HF, quartz fibers of two different lengths are embedded in the steel, and are read separately.
The calorimeter is thus functionally subdivided into two longitudinal segments\,(not shown in Fig.~\ref{fig:hcal_layout}).
Long fibers\,(165\,cm $\approx 10$ interaction lengths) measure the total signal coming from the full material length, whereas short
fibers measure the energy deposition after 22\,cm of steel.
This allows the separation of showers generated by electrons and
photons from those generated by hadrons.
The photodetectors for HF are eight-stage
photomultipliers\,(PMT) with a borosilicate glass window,
which register the Cherenkov light produced by the charged shower particles
in the quartz fibers.

The signals from both types of photodetectors
are read by a custom-designed  chip,  which performs charge integration and encoding\,(QIE)~\cite{bib:2}. The QIE is a non-linear
multi-range analog to digital converter\,(ADC) that provides approximately constant fractional precision
over a wide dynamic range. For each channel, the measured ADC value is converted into
a charge measured  in femtocoulomb\,(fC),  using a conversion factor that was determined in the laboratory  by charge injection into the QIE.

Each subsystem is equipped with laser and light emitting diodes\,(LED), for monitoring and calibration.
In addition, the response of each HB and HE scintillator tile of each layer and each HF fiber can be measured using radioactive
sources mounted on the tip of a moving wire~\cite{bib:source}.

The Electromagnetic Calorimeter\,(ECAL),
made of lead tungstate crystals, is situated in front of HB and HE.
The response and resolution of the
CMS calorimeter system
depends on both the ECAL and the HCAL, as most particles start showering in the ECAL. The
hadronic to electromagnetic\,($\pi/e$) response ratios of the ECAL and the HCAL are different:
typical values of $\pi/e$ for the ECAL
and the HCAL are 0.61 and 0.80 at 10\,GeV/$c$, and 0.75 and 0.85 at 100\,GeV/$c$,  respectively\,\cite{bib:4}.
The ECAL and the HCAL fraction of the energy deposited in each calorimeter varies non-linearly with energy and,
 as a result, the raw energy measurements require substantial corrections.
Results from test beam studies are reported in~\cite{bib:1,bib:2,bib:3,bib:4}.

Parts of the various HCAL subsystems were exposed to beams of electrons, pions,
protons
and muons, to measure their
 characteristics and to obtain a reference
calibration. An ECAL module was also included in the test beam setup.
The hadronic energy resolution of the barrel HCAL and
ECAL combination being  parameterized as
${\sigma}/{E}={a}/{\sqrt{E}}\oplus b$, where $a$ corresponds to a stochastic term and $b$  to a  constant term,
one measures  $a=0.847\pm0.016$\,GeV$^\frac{1}{2}$
and $b=0.074\pm0.008$\,\cite{bib:2},  with $E$ measured in GeV.
The energy resolution in the endcaps is similar to that in the
barrel. The corresponding values for HF are $a=1.98$\,GeV$^\frac{1}{2}$
and $b=0.09$\,\cite{bib:1}.
Since the forward jets typically have very high energies,
the stochastic term may be higher for HF than for the other
calorimeters, while still providing the required energy resolution~\cite{bib:1}.

\section{Impact of the Magnetic Field on the HCAL     Response}

\label{sec:magnet}

About 270 million cosmic ray muons were recorded during CRAFT with the magnetic field on.
Typically, a muon event is triggered by the muon chambers and
the muon momentum is measured by the central tracker and muon chambers.  Clean signals were measured in most of the HCAL towers.
The cosmic ray muons are used to measure the HPD gain
to an accuracy of about  5\% for HB and HO towers, and 10\% for HE.
Calibration corrections based on the study of the  cosmic ray muon samples
have been derived and applied for HB and HO,  as discussed in  Section~\ref{sec:validation}.

Brightening corrections,  of about 8-10\%,
applied to HB and HE calibration parameters to take into account the effects of
the magnetic field were already
obtained during MTCC, where it was possible to perform precise measurements
using radioactive sources~\cite{bib:det}.
Cosmic ray  muon data from CRUZET and CRAFT allowed to test these corrections and to
study the impact of the strong magnetic field in other parts of the HCAL system.

\begin{figure}[ht]
  \begin{center}
\includegraphics[width=0.45\textwidth]{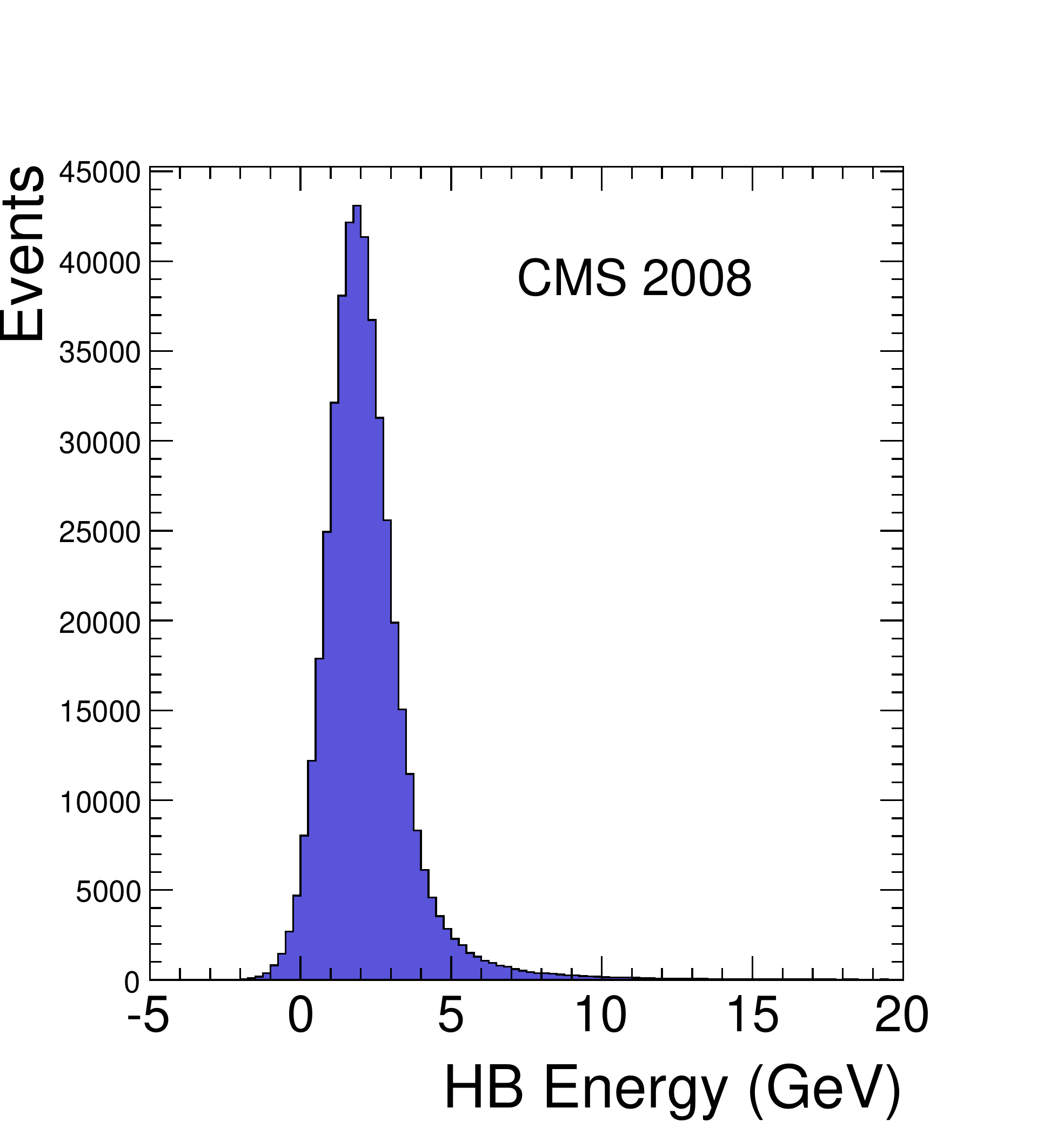}\hspace{0.25cm}\includegraphics[width=0.475\textwidth]{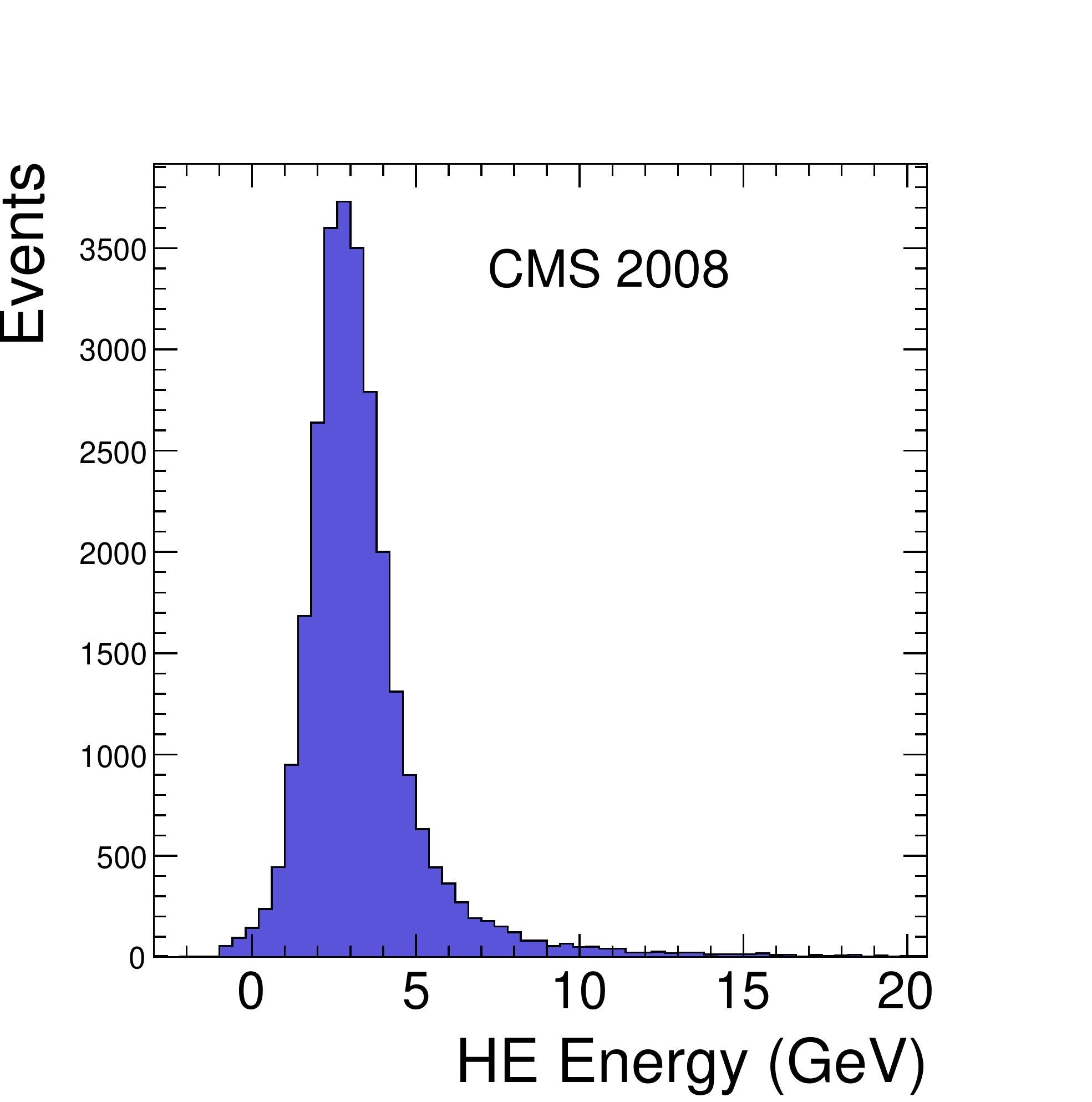}
    \hspace{1cm}
    \caption{Cosmic ray  muon energy deposition measured in HB\,(left) and HE\,(right) with $B = 3.8$~T for muon  momenta
above 7\,GeV/$c$. HB and HE data samples consist of 450 and 27 thousand events,
 respectively.}
    \label{fig:muon_signals_hb_he}
  \end{center}
\end{figure}

\subsection{The Barrel and Endcap Hadron Calorimeters}

A single muon traversing the HCAL deposits a few GeV in HB and HE.  Figure~\ref{fig:muon_signals_hb_he}
shows the energy deposited in HB and HE by muons with momenta above 7\,GeV/$c$, during CRAFT with the 3.8~T magnetic field.
The means are consistent with
the muon minimum ionizing signals measured for segments of the calorimeters in test beams~\cite{bib:2,bib:4},
without magnetic field,
for a wide momentum range 
after taking into account magnetic field effects, correcting the signal for the muon path length
and normalizing to the thickness of HB at $\eta=0$.  The width of the distributions of the cosmic ray  muons
are
broadened by the momentum spread
of cosmic ray muons, since
the energy deposition increases with momentum due to the relativistic rise. The measurement of the
energy deposition by cosmic
ray muons is  performed  using either the electron or the pion calibration of the calorimeter.
As an example,
the mean energy deposition of 150\,GeV/$c$ muons is calculated  to be 2.4\,GeV\,(2.8\,GeV) if the electron\,(pion) calibration
is used.
For Fig.~\ref{fig:muon_signals_hb_he}, the pion calibration was used. The average values of the energy deposition is lower
than 2.8\,GeV because the average muon momenta is well below 100\,GeV/$c$\,(see momentum spectrum given in 
Fig.~\ref{fig:dedx}).

It is known\,\cite{bright1,bright2,bright3,bright4,bib:5},
 that the presence of a magnetic field increases the signal measured in
scintillator based calorimeters.
There are three sources contributing to this effect.
The first of them  is due to a  chemical effect that  causes an intrinsic brightness of the scintillator.
Brightening is due to the polymer base of the scintillator and saturates at about 2\,T.
The  maximum signal increase due to brightness  ranges
 between 5\% and 8\%, depending  on the composition of the plastic \cite{bib:5}. The field
value at
HB and HE is  3.8\,T, which  is well above the saturation value.
%
The second source of signal increase is the curling of knock-on electrons 
in the magnetic field, which results in an increased path length of the 
electrons,
and enhances the signal by 1-2\%.
Therefore, the signal increase
depends on the nature of the particle impinging on the
calorimeter, on the geometry and on the strength of the magnetic field.  For this effect, the specific
orientation of the magnetic field relative to the absorber plates is thus  important and is different for HB and HE.
At the LHC, the directions of particles emitted from the IP are essentially normal to the scintillator plane,
the magnetic field direction is parallel to the scintillator plane in HB and perpendicular in HE.
Previous measurements\,\cite{bib:5}
and this study showed that the light yield increase due to the magnetic field is the same
for muons and for particles from  radioactive sources,
while  another complementary  study~\cite{bib:5} showed that the effect is different
for pions and electrons,  as electrons produce more
knock-on electrons than pions.
There is a third effect due to the HPD geometry. When a muon passes through
a single tower, all  the photons impinge on one single HPD pixel. However, the electrons from the
photocathode spread out  and up to 2\% of the signal is measured in neighboring pixels,
a phenomenon called cross talk. The magnetic field helps in reducing this  cross talk, but
any misalignment of the HPD axis with respect to the magnetic field axis increases the cross talk.

\begin{figure}[hbtp]
  \begin{center}
    \includegraphics[height=0.25\textheight,width=0.42\textwidth]{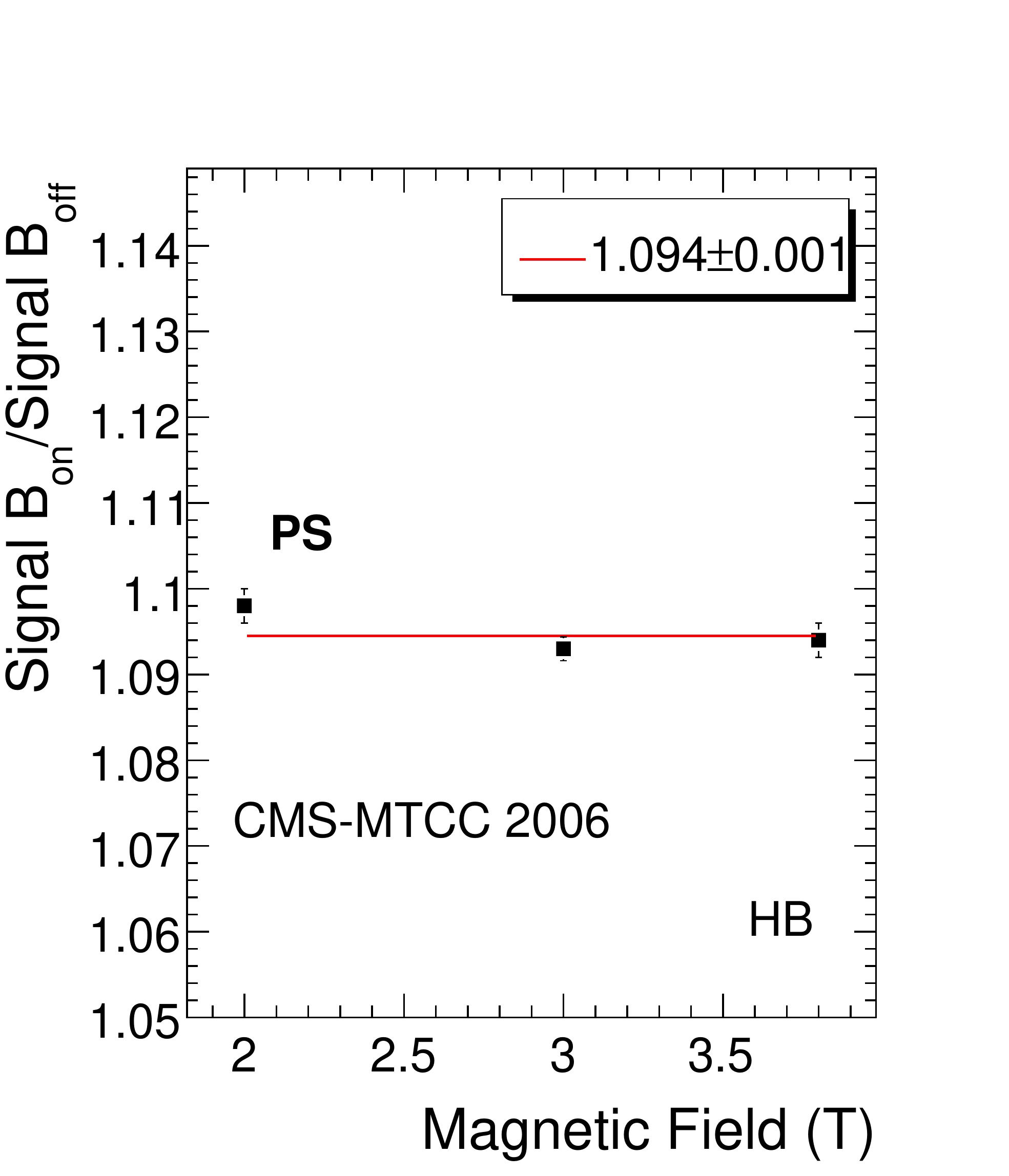}\hspace{1cm}\includegraphics[height=0.25\textheight,width=0.425\textwidth]{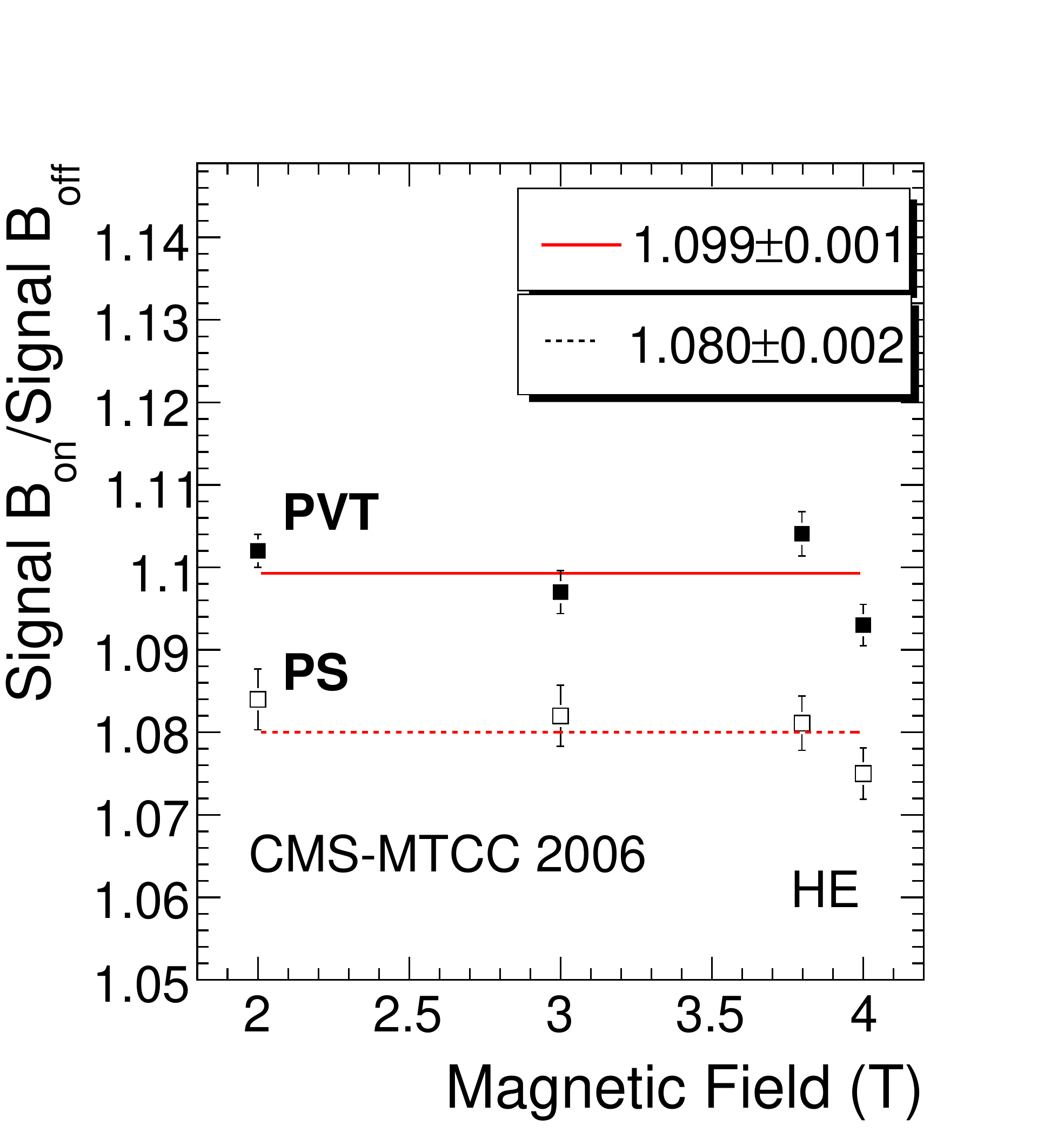}
    \hspace{1cm}
    \caption{Ratio of the scintillator signals in a single layer  with magnetic field on and off, measured using a radioactive source during MTCC.
The left figure is
 for polystyrene\,(PS) in HB  and the right figure is PVT\,(filled
squares) and PS\,(open squares) in HE. The horizontal lines represent the averages of each data set.
The difference for PS between HB and HE is due to the curling of the knock-on electrons and the magnetic field direction.
}

    \label{fig:magnet_effect_hb_he}
  \end{center}
\end{figure}

The magnetic field effect on the HCAL was observed by comparing the measurements
performed with a   moving radioactive wire
source, with field on and with field off during the MTCC. The inners layer\,(Layer~0) of HB and HE and all layers
of HO are made of 1\,cm
thick polymer polyvinyl toluene\,(PVT);  all other layers of HB and HE are made of 0.4\,cm
thick polymer polystyrene\,(PS).
Figure~\ref{fig:magnet_effect_hb_he} shows the measurements
performed  at several magnetic field values for a typical polystyrene layer of HB\,(left) and for two layers of HE\,(right),
one  PVT  and one PS.
 The PVT layer of HB was not installed during MTCC.
The magnetic field effect on PS of 9.4\%
measured in HB during MTCC agrees well with the
expected value. The 2\% difference in brightening between PVT and
PS, shown in Fig.~\ref{fig:magnet_effect_hb_he}\,(right), was also observed in an independent
study~\cite{bib:5}. For HE, a signal enhancement of about 6\% is due to the increased brighteness of the
polymer polystyrene, and the rest is due to the path length of the knock-on electrons.
The comparison of CRAFT and CRUZET data reveals a net signal enhancement at 3.8\,T of $1.090 \pm 0.005$ for HB and
$1.084 \pm 0.013$ for HE, consistent with the MTCC measurements of  $1.094 \pm 0.001$ for HB
and $1.080 \pm 0.002$ for HE.

\subsection{The Outer Hadron Calorimeter}
\label{sec:HO-mag}

\begin{figure}[hbtp]
  \begin{center}
    \includegraphics[height=0.245\textheight, width=0.38\textwidth]{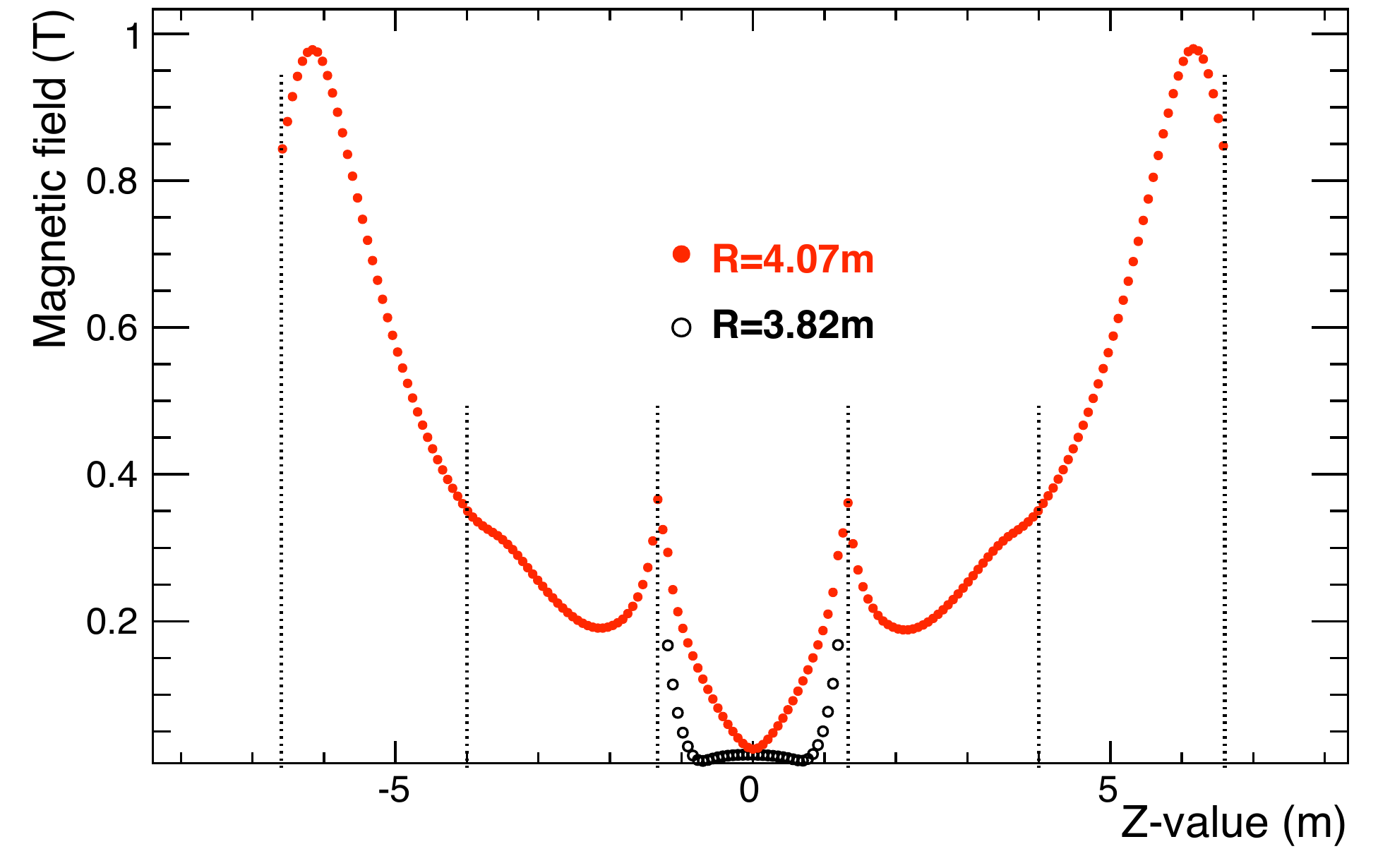}\hspace{1cm}\includegraphics[height=0.25\textheight,width=0.45\textwidth]{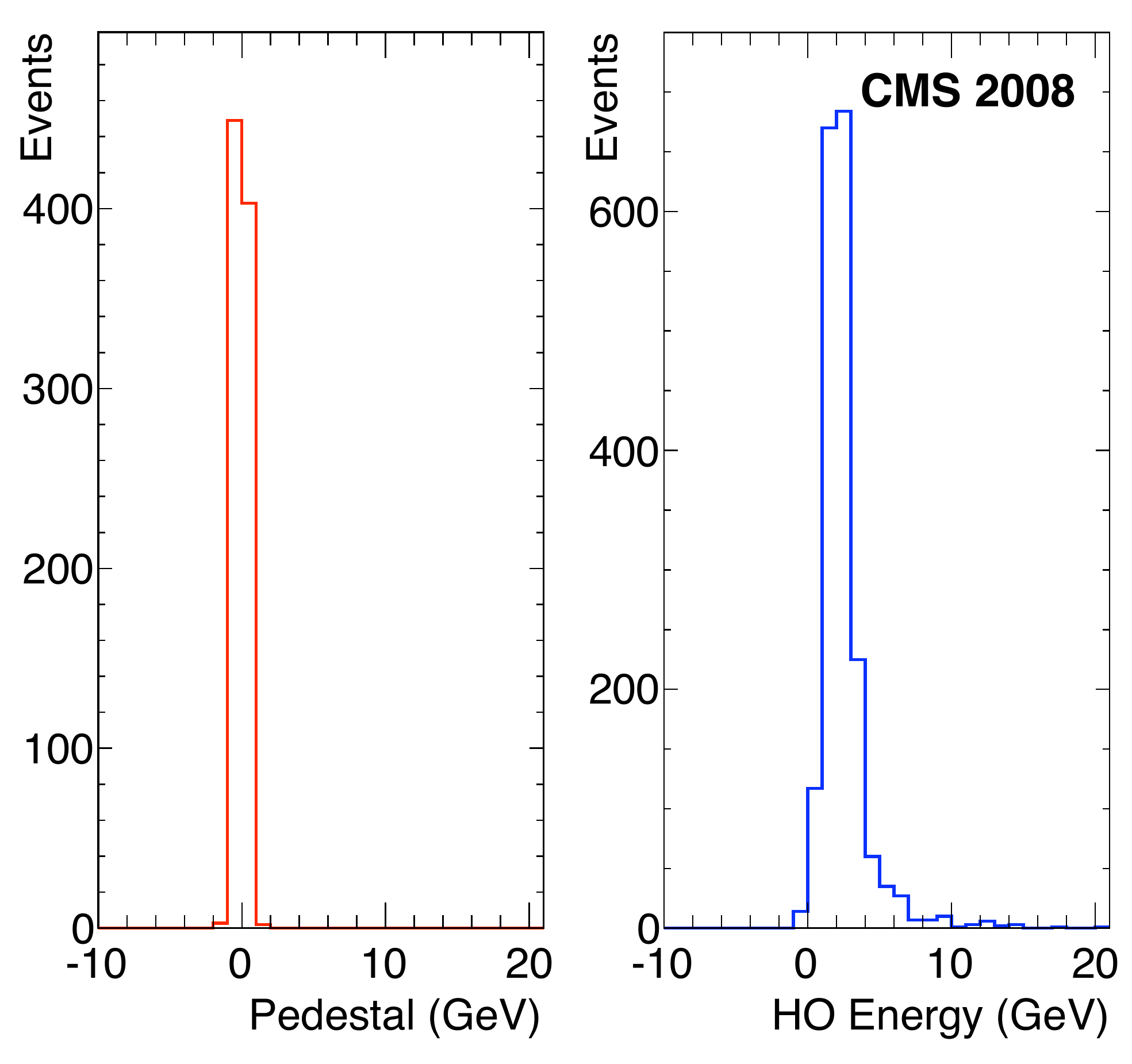}
    \hspace{1cm}
     \caption{Left: Magnetic field strength at the location of the HO scintillators, as  calculated
with the TOSCA program\,\cite{bib:tosca}. The vertical lines indicate the ring borders. Ring~0
has two scintillator layers, at radii 3.82\,m and 4.07\,m, respectively.
Right: Pedestal and cosmic ray muon signal in Ring~0 of HO.}    \label{fig:muon_signals_hO}
  \end{center}
\end{figure}

All scintillator layers in HO are made of 1\,cm thick PVT.
Ring~0 has        two scintillator layers, while
Rings~1 and 2 have  a single scintillator.
Figure~\ref{fig:muon_signals_hO}\,(left) shows the field strength at the position where the scintillators are located.
The first scintillator layer of Ring~0, at a radius of 3.82\,m, is located in a region with a very
small magnetic field. The signals in the two layers in Ring~0 are added optically and the brightening effect is the average of
the effects in the two scintillator layers.
The other scintillator layers are
at a radial  distance of 4.07\,m from the beam line,
where the magnetic field is more important. The scintillator brightening for Ring~0 is about 1\%, and for
Rings~1 and 2 it varies between 2\% and 4\%.  This is to be compared to the 8-10\% effect observed in HB and HE above 2\,T.

Cosmic ray muon signals are observed in all rings of HO.
In Ring~0, with two scintillator layers, the signals are twice as large as in Rings~1 and 2.
Figure~\ref{fig:muon_signals_hO}\,(right) shows the distributions of the pedestal and
the pedestal subtracted  muon signal in Ring~0.

\subsection{The Forward Hadron Calorimeter}

The magnetic field strength
in the HF region is significantly reduced compared to
that in the HB and HE regions, allowing the use of conventional PMTs.

During CRAFT, the magnetic field at the PMT locations was measured to be less than 0.03\,T,
and proper  PMT operation was verified at full magnet current.  Each PMT is shielded with individual
tubes made of a
nickel-iron alloy and of soft iron.  HF was fully operational and stable during CRAFT and the LHC startup.

In the  CMS cavern, the number of cosmic ray muons decreases significantly with increasing incidence angle
relative to the vertical direction.  Therefore, it was
 not possible to validate the calibration of HF channels with the
CRAFT data,  due to the lack of muons in the horizontal direction.

The LHC startup data from September 2008, however, proved useful in validating the calibration constants, 
and will be  discussed in
Section~\ref{sec:validation}.  Occasional abnormally large signals were found in individual HF channels during the LHC startup.
The signals were found to be due to muons striking directly the window faces of individual PMTs.  Algorithms
for rejecting such abnormal signals, by comparing the energies reconstructed from 
long and short fibers with the same ($\eta,\phi$) values, were subsequently developed  based on these data~\cite{bib:7}. 
The    LHC beam halo data show that the current algorithms can reduce the rate 
of such abnormal signals by a factor of one thousand.

\section{Optimization of the HCAL Operation   using CRAFT Data}
\label{sec:char}
This section describes the information obtained from CRAFT about noise levels, detector stability and optimization
of hardware settings, like high voltage and zero-suppression parameters,
which could influence the HCAL energy resolution.

The HCAL has a total of 2592 channels in HB, 2592 in HE,
1728 in HF and 2160 in HO.  At the beginning of CRAFT,  99.3\%  of HB, HE and HF,
and 95.5\% of HO channels were operational.  About 75\% of the non-operating channels had
photodetector\,(HPD or  PMT) problems and the rest had    electronics
failures of various types.

Several HPDs, about 10\% of the total, functioned poorly when the
magnetic field reached full strength.
The high voltage for those channels   was either reduced by 1.5\,kV,
leading to a $\sim30\%$ reduction in gain, or was completely turned off.
After this action was taken, 87\% of the HCAL was operating at the nominal high
voltage of 7.5\,kV both for the barrel and the endcaps.
No evidence of further degradation of HPD functioning was observed during CRAFT.
In HF,
which employs PMTs, the high voltage was set at 1.25\,kV  as originally planned.

The problematic HPDs were replaced after CRAFT.
As of September 2009,  all the channels in the barrel, endcap and forward regions are
operational at full magnetic field, while the number of non-operational channels in HO
is at the level of 2\%.
As a precaution, the nominal high voltage during LHC collisions will be set 0.5\,kV lower
than in CRAFT for the barrel and the endcaps,  at a cost  of a  12\% reduction in gain.
The impact on jet energy resolution is small, of the order of 5\%.

\subsection{Performance of Photodetectors and Impact on the Trigger}
\begin{figure}[hbtp]
  \begin{center}
    \includegraphics[width=0.45\textwidth]{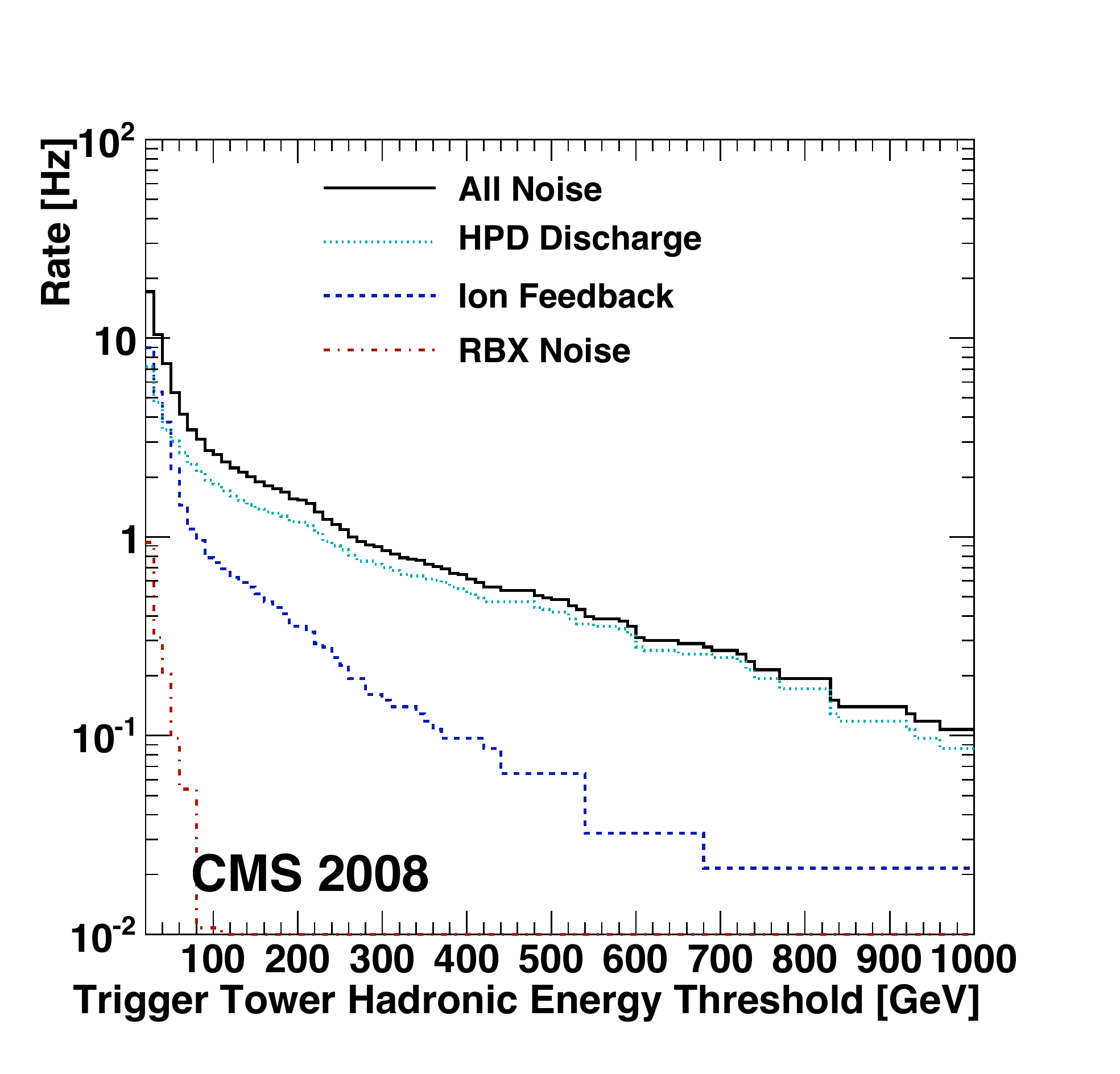}
    \hspace{1cm}
    \caption{Total noise rate for around 4000 trigger towers as a function of the
energy threshold,
 during CRAFT with the magnet at full field.  A minimum energy of 20\,GeV was required in at least   one
HB or HE readout box\,(RBX).
About 280 HPDs, corresponding to  5040 channels of
HB and HE, are included in the analysis; HPDs that were replaced in the beginning of 2009 are excluded.}
    \label{fig:hpd_noise}
  \end{center}
\end{figure}

During CRAFT, the rate of HPD noise replicating an
 energy deposit of 20\,GeV or more in a single channel
was below 0.2\,Hz~\cite{bib:7}, after the malfunctioning HPDs were excluded.
Figure~\ref{fig:hpd_noise} shows the measured total noise rate for all trigger towers in the barrel and the
endcaps,
as a function of the energy threshold.
The signals of the longitudinal segments of a tower are added to form a trigger
tower in HB, in HB-HE interface region and in HE\,(Fig.~\ref{fig:hcal_layout}).
Three sources contribute to this noise rate~\cite{bib:7}: (a)  the  ion feedback inside
the HPD (two to three neighboring HPD pixels with energy greater than
1\,GeV), which affects mainly single channels, (b) the noise from the whole HPD,
which affects 18 channels simultaneously
(all HPD pixels with energy greater than 1\,GeV), and (c) coherent noise in a readout box\,(RBX),
which affects a full set of 72 channels (all channels with energy greater than 1\,GeV); one readout box includes
the signals from four HPDs.  The ion feedback is  predominantly caused
by a thermally emitted electron that ionizes the residual gas inside the HPD
acceleration gap. The ion is accelerated back to the cathode and liberates
further electrons, causing a signal equivalent to many photoelectrons.
Misalignment of the HPD axis with respect to the axis of the solenoidal
field can reduce the voltage of the HPD above  which electrical discharges can occur.
This can lead to an avalanche of secondary electrons,
producing significant energy deposition in a large number of
channels within an HPD.  This is the main noise contribution above 100~GeV and
is referred to as ``HPD discharge''.
The cause of the readout box coherent noise is not yet understood, but measurements
show that it does not have an impact on the trigger and can be easily
identified  during  the  event reconstruction.

The HPD noise, with or without  magnetic field, has a small   impact on the
total trigger rate.
For example, without  magnetic field and with a threshold of 10~GeV, the HCAL
self-trigger rate is 130\,Hz; at full field the rate increases by 40\,Hz.
This is to be compared with the 100\,kHz trigger rate expected during
collisions.
Furthermore, the rate decreases rapidly as a function of the applied threshold,
as seen in Fig.~\ref{fig:hpd_noise}.

Special triggers to monitor the  HPD noise rate, offline
software to identify the corresponding atypical  signals at the event
reconstruction level~\cite{bib:7}, and the corresponding
simulation tools have been developed and   are available.
Initial tests on data show that the algorithms for identifying noise and 
problematic channels correctly mark both persistent and sporadic known 
problems with high efficiency.

\subsection{Performance of Readout Electronics and Impact on Energy Thresholds}

\begin{figure}[hbtp]
  \begin{center}
    \includegraphics[width=0.425\textwidth]{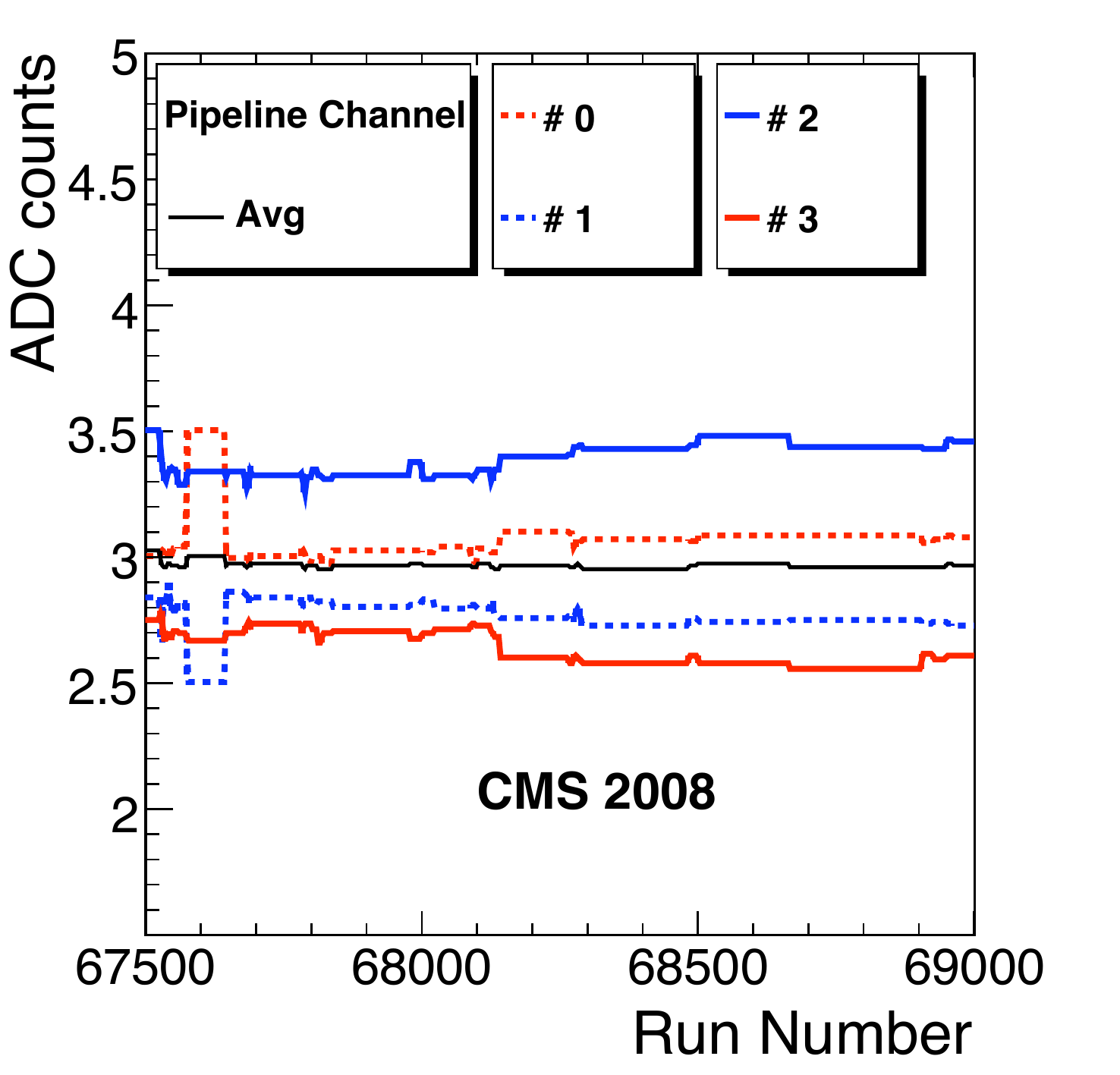}\hspace{1cm}\includegraphics[height=0.2775\textheight,width=0.425\textwidth]{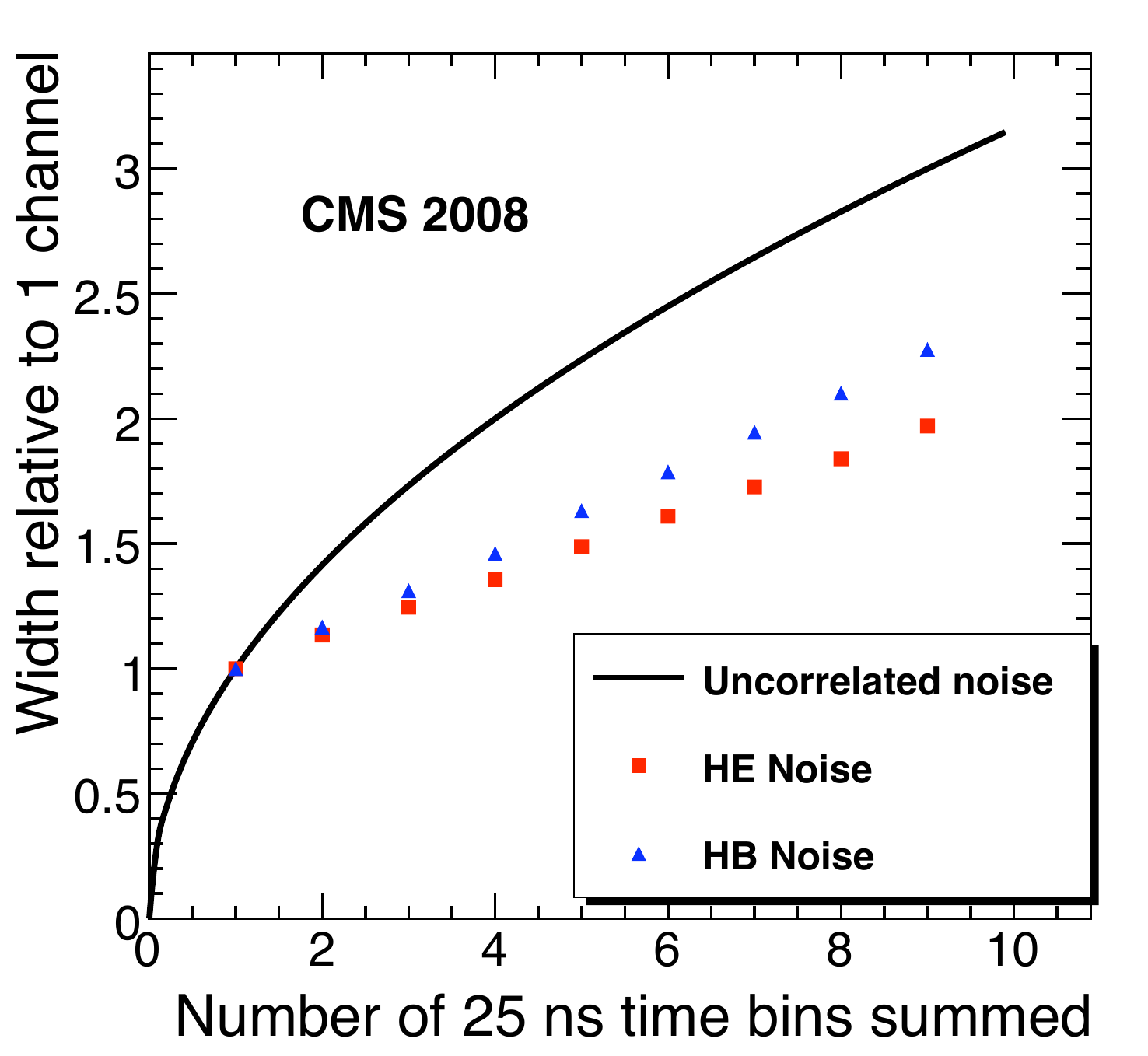}
    \hspace{1cm}
    \caption{Left: The four pedestal values corresponding to the four 25\,ns channels
and their average, for one HB calorimeter tower over a period of 10~days. Right:
The measured pedestal width in HB and HE when averaging over several\,(1 to 9) time
slices\,(TS). The solid line represents the expected noise\,($\propto\sqrt{n}$) in the absence of correlations.
The signal reconstruction will sum four TS at low luminosity and two TS at high luminosity.
The single channel noise at 7.5\,kV for 1\,TS is 140 and 200\,MeV for HB and HE, respectively.}    \label{fig:electronic_noise}
  \end{center}
\end{figure}

The electronic noise and the pedestals in the QIE chips were found to be stable during CRAFT.
The QIE chips read the signal over four time intervals of 25\,ns each, using four pipeline-channels
(numbered 0 to 3), and output four ADC values; the cycle repeats every 100\,ns.
For each QIE one measures  four  pedestal values of a few ADC counts each (one for each 25\,ns time interval),
which are subtracted from the corresponding signal.
The left plot in Fig.~\ref{fig:electronic_noise} shows the values of the four pedestals and
their averages for a particular HB tower, over a period of ten days. There is a small variation
of the individual pipeline-channel  pedestals over time, with a strong anti-correlation
among them.  The average of the four pedestals does not change over this time period.

The four pedestal
values are correlated due to the integration of the pulse shaper,
as demonstrated in the right plot of Fig.~\ref{fig:electronic_noise}. This figure
shows the observed noise averaged over a variable number $n$\,(from 1 to 9)  of 25\,ns time periods, compared
to the increase as $\sqrt{n}$ expected in the absence of correlations.
These noise correlations have been included in the HCAL simulation.

The width of the pedestal distribution allows one  to estimate the QIE electronic noise.
For example,  the measured electronic noise during CRAFT in a {\it single} time slice corresponds
to an equivalent energy in a tower of 140\,MeV for HB, 200\,MeV for HE, 280\,MeV
for HO and 370\,MeV for HF\,(the larger HF noise remains small compared to the energy
corresponding to a  single photoelectron, 4\,GeV,
and to the typical energies of particles impinging HF).
The increase when integrating over more than one
time slice is modest, as can be inferred from Fig.~\ref{fig:electronic_noise}\,(right).

The HCAL information obtained from a tower is not recorded if the signal is below a
threshold, that is, the HCAL data are ``zero-suppressed''.
The information of a tower is kept only if
the sum of ADC counts in two adjacent time slices is larger than 2~counts
above pedestal for HB,  3~counts for HE/HO,
and 4~counts for HF. After taking into account the final high voltage settings planned for
collision data, these thresholds
correspond to $\sim$400\,MeV for HB,    $\sim$700\,MeV for HE,
$\sim$1\,GeV for Ring~0 of HO, and
$\sim$1\,GeV for HF.  These threshold  values were selected such that the HCAL occupancy
is reduced to about 20\% at the nominal LHC luminosity.
Since some towers, which contain little energy, are eliminated by zero-suppression, this can
reduce the measurement accuracy.
However, since the energy thresholds for zero-suppression are well below the energy threshold
applied in the jet reconstruction, and since a shower is distributed over many towers, zero-suppression
will have little effect on jet and single high energy hadron energy  measurements.
It can, however, affect the HCAL energy measurement for low-momentum muons\,($p_{\mu}<10$\,GeV/$c$)
traversing a single tower
and all muons traversing multiple towers.

\section{Calibrations using Cosmic Ray Muons and LHC Beams}
\label{sec:validation}

The calibration of the HCAL has been performed using several techniques. First, test beam setups were used to determine,
for parts of the HCAL subsystems, the detector response to radioactive sources and to electrons, muons and hadrons
of known momenta.
Then, the intercalibrations of all channels in HB, HE, and HF were performed with radioactive sources
mounted on moving wires, exactly as was done in the test beam setups.
These source measurements also propagate the absolute energy scale for each subsystem.
The energy scale for the reference Co$^{60}$ source was made  with respect to 50\,GeV/$c$ pions in the case
of HB and HE, while for HF 100\,GeV/$c$
electrons were  used.



A total of 31 beam splash events from the beam arriving from the $+z$ side and 35 events from the $-z$
side were recorded during the 2008 LHC commissioning exercise, with the CMS magnetic field off.
In these events, muons from secondary decays penetrated the detector.
Since the scintillator planes of HB and HO are parallel
to the beam line, the muons from  splash events traverse the full length of the scintillators
giving a very large signal.
The  splash events could not be used to study the absolute energy scale calibration  because their
energy  and muon flux is not accurately known.
However, these events  were         helpful for measuring the relative scale between $\phi$ sectors and between
the $\pm z$ halves of HB and the two HE endcaps.

The initial calibration obtained from measurements with radioactive sources was  tested, and in some cases
improved, using cosmic ray muons and the beam splash events
for a significant fraction of HB channels.
 However,
since cosmic ray muons in the CMS cavern are essentially vertical, HB $\phi$ sectors in the horizontal plane near
$\phi=0^\circ$\,(sectors 71, 72, 1 to 4) and near $\phi=180^\circ$\,(sectors 33 to 38)
were not included in this study, while in HO all channels were calibrated using cosmic ray muons.
All channels in HE and HF were mainly tested with beam splash events.

To ensure good energy measurement, only events collected
without zero-suppression are used in the calibration studies.


\begin{figure}[t]
  \begin{center}
    \includegraphics[height=0.30\textheight,width=0.425\textwidth]
{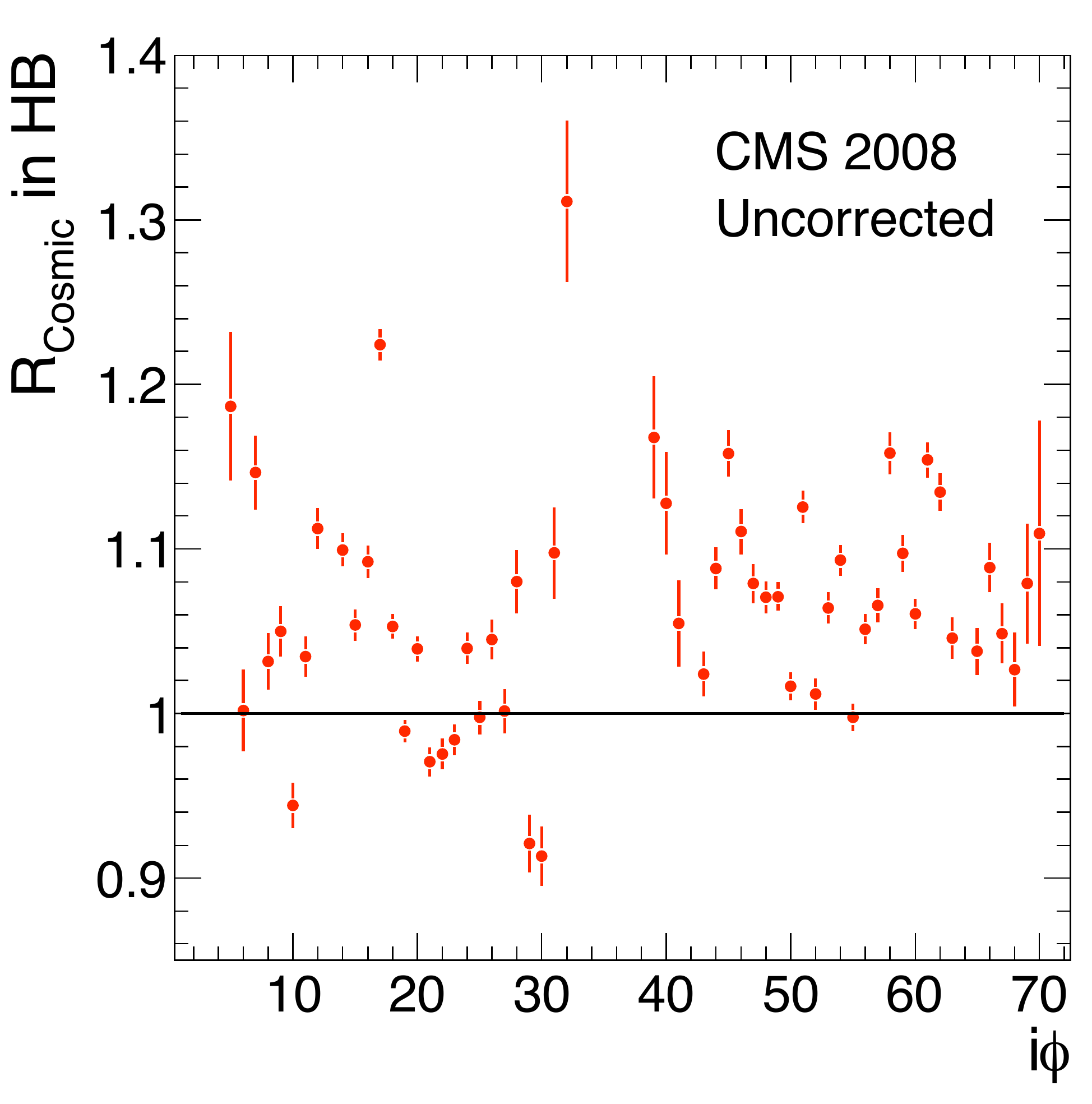}
\hspace{1cm}
\includegraphics[width=0.435\textwidth]
{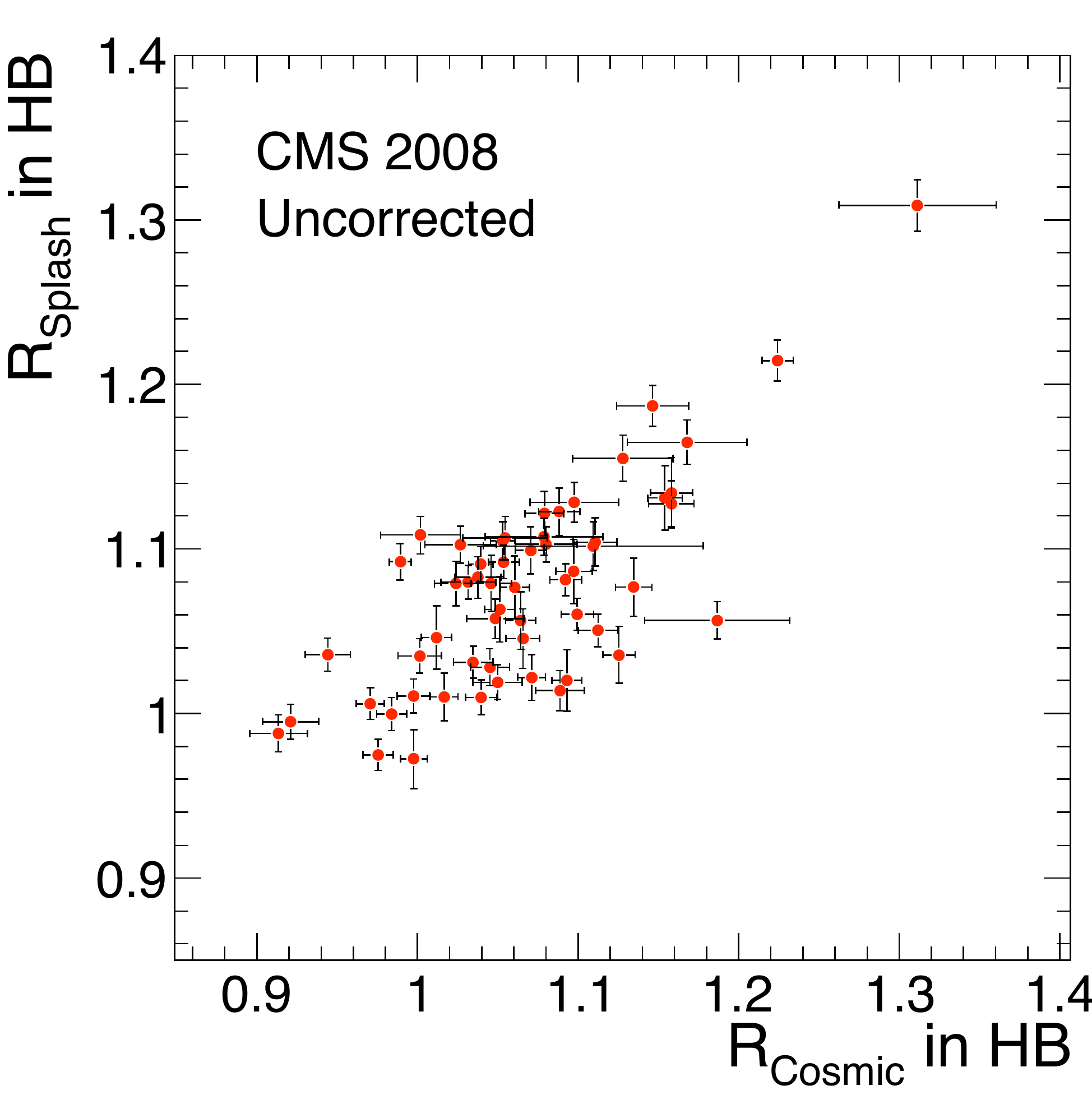}
    \hspace{1cm}
    \caption{
Left:
Ratio, for cosmic ray muons, of the average signal amplitudes in the two halves\,($+z$ and $-z$)
of HB,  as a function of the $\phi$ sector index, $i\phi$, 
after correcting the average signal for the muon path length
and normalizing to the thickness of HB at $\eta=0$.
Right:
Correlation between cosmic ray muon and splash muon signals, for the ratio
of the average signal in the two halves of HB.
The splash muon signals were corrected for the asymmetry in the muon flux by comparing the
events from the beam arriving from the $+z$ side and the $-z$ side.
}
    \label{fig:hb_correction}
  \end{center}
\end{figure}

\subsection{The Barrel and Endcap Hadron Calorimeters}

All HB and HE scintillator tiles had their response measured in-situ
using either 1.5 or 5\,mCi Co$^{60}$ sources. The averages of the signals in the 17 layers
of HB and HE were used to equalize the response of all HB and HE towers.
The absolute energy scale was set using the response to 50~GeV/$c$ charged pions
of the parts of the detector exposed  to the  test beam studies.

%

\begin{figure}[hbtp]
  \begin{center}
    \includegraphics[height=0.30\textheight,width=0.425\textwidth]{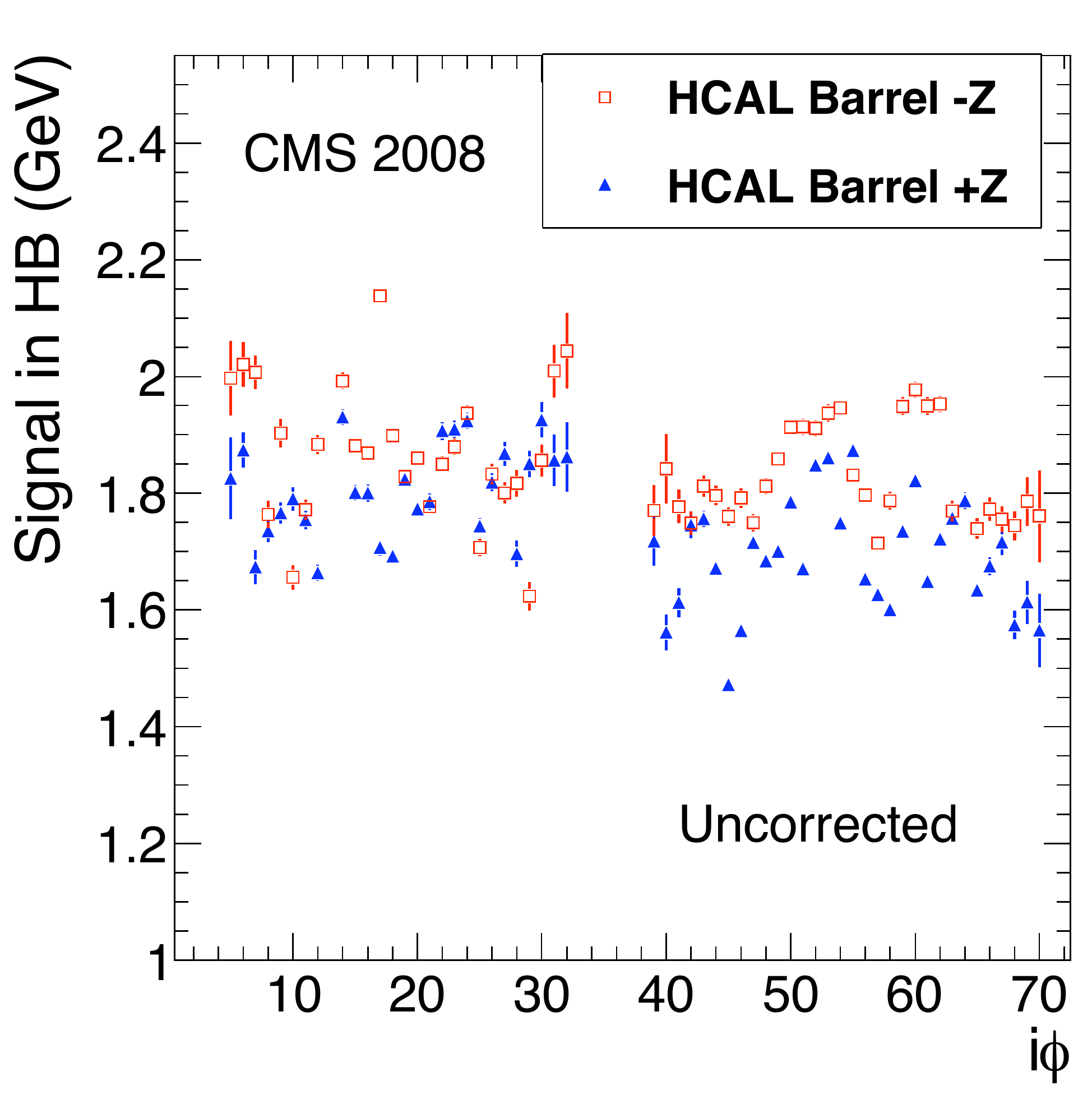}\hspace{1cm}\includegraphics[width=0.435\textwidth]{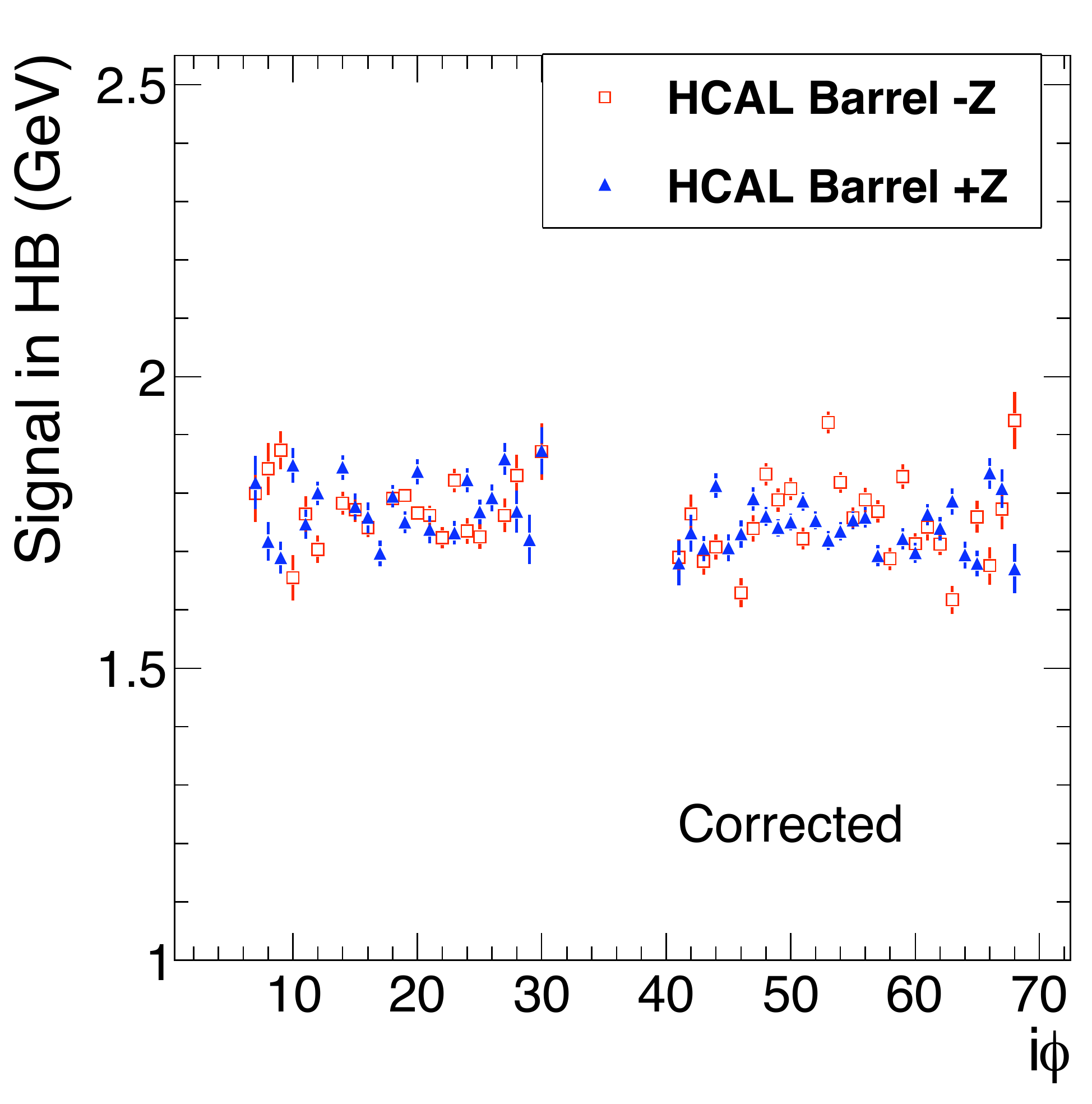}
    \hspace{1cm}
    \caption{Cosmic ray muon signal in HB averaged over all $\eta$ values
         for each $\phi$ sector, $i\phi$,  before\,(left) corrections for a sample of $\sim$400
         thousand events
         and after\,(right) corrections for sample of $\sim$80 thousand  events.
         The triangles\,(open squares) represent the results for the
         $+z$\,($-z$) side of the detector.
         The mean and RMS values for the muon signals for four HB regions   are presented in
Table~\ref{tab:sum_corr_hb}.
}
    \label{fig:hb_validation}
  \end{center}
\end{figure}

\begin{table}[hb]
\begin{center}
\caption{Mean and RMS of the muon energy distributions measured in four regions of
 HB, before and after correcting
the calibration constants using cosmic ray muon data information.
\label{tab:sum_corr_hb}}
\begin{tabular}{|l|c|c|c|c|}
\hline
Detector region& Mean before  & RMS  before  & Mean after & RMS after \\
region   & correction\,(GeV) & correction\,(GeV) & correction\,(GeV) & correction\,(GeV)\\
\hline
HB-: $i\phi$ 5-32  & 1.87 $\pm$ 0.03 & 0.12 & 1.78 $\pm$ 0.01 & 0.05\\
HB-: $i\phi$ 39-70 & 1.83 $\pm$ 0.02 & 0.08 & 1.75 $\pm$ 0.02 & 0.07\\
HB+: $i\phi$ 5-32  & 1.81 $\pm$ 0.02 & 0.08 & 1.78 $\pm$ 0.01 & 0.05\\
HB+: $i\phi$ 39-70 & 1.70 $\pm$ 0.02 & 0.09 & 1.74 $\pm$ 0.01 & 0.04\\
\hline
\end{tabular}
\end{center}
\end{table}

The cosmic  ray muons and splash events
were used to check the initial calibration conditions.
The results from both methods point to non-uniformities in HB calorimeter response
in different $\phi$ sectors, and to a difference in the energy scale between the two halves of HB.
Figure~\ref{fig:hb_correction}\,(left) shows the ratio of the muon
signal amplitudes  in HB for $z<0$ and $z>0$ for cosmic ray muon events, $R_{Cosmic}$,
as a function of the  $\phi$ sector index, $i\phi$. While
Fig.~\ref{fig:hb_correction}\,(right) shows that measurements for this ratio based on
cosmic ray muons and beam splash events, $R_{Splash}$, have a 75\% correlation.
The average ratio deviates from unity by $\sim$7\% with  a
sector-by-sector spread  of about $10\%$\,(see Table~\ref{tab:sum_corr_hb}).
The source of these differences is not understood at this time.

The results from the cosmic ray  muons were used to correct the calibration of the
various HPDs, with one single parameter applied to all channels contained in a given $\phi$
sector of one of the two $\pm z$ halves of HB. Only muons traversing a single $\phi$ sector and less than five
$\eta$ channels were used in the analysis.
About half a million CRAFT muons with  momentum below 100\,GeV/$c$
remain after the full event selection.
Sector~13 of HB, where an HPD was  replaced before CRAFT, was excluded from these studies.
Figure~\ref{fig:hb_validation} shows the peak value of the distribution of the average  energy deposited in
HB by cosmic  ray muons, averaged over all
$\eta$ values  of a given $\phi$ sector for two independent data samples. The plot on the left
shows the results for the data used to determine the corrections, while the plot on the
right shows  the results after the corrections are applied to a   smaller but independent data sample.
Table~\ref{tab:sum_corr_hb} presents the corresponding results for four HB regions.
After corrections,  the signal spread is  smaller and the differences between the  mean values for the four
regions   of HB is significantly reduced, reflecting the improvement of the HPD intercalibration.
The mean muon energy measured in this analysis is smaller than the one
shown in Fig.~\ref{fig:muon_signals_hb_he} because
only tracks with a momentum below 100\,GeV/$c$ were used.

\begin{figure}[hbtp]
  \begin{center}
    \includegraphics[height=0.285\textheight,width=0.425\textwidth]
    {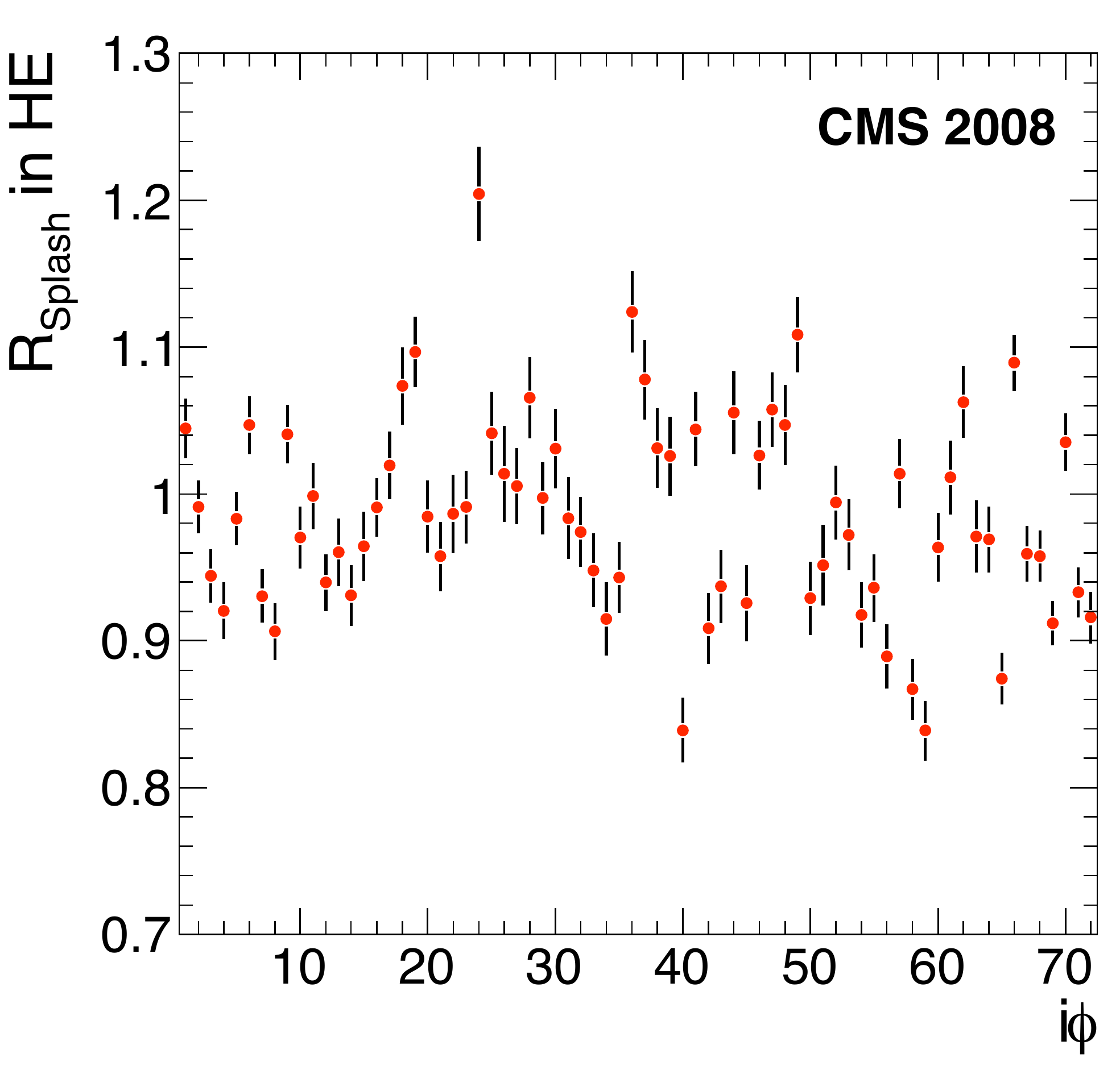}\hspace{1cm}\includegraphics[width=0.435\textwidth]{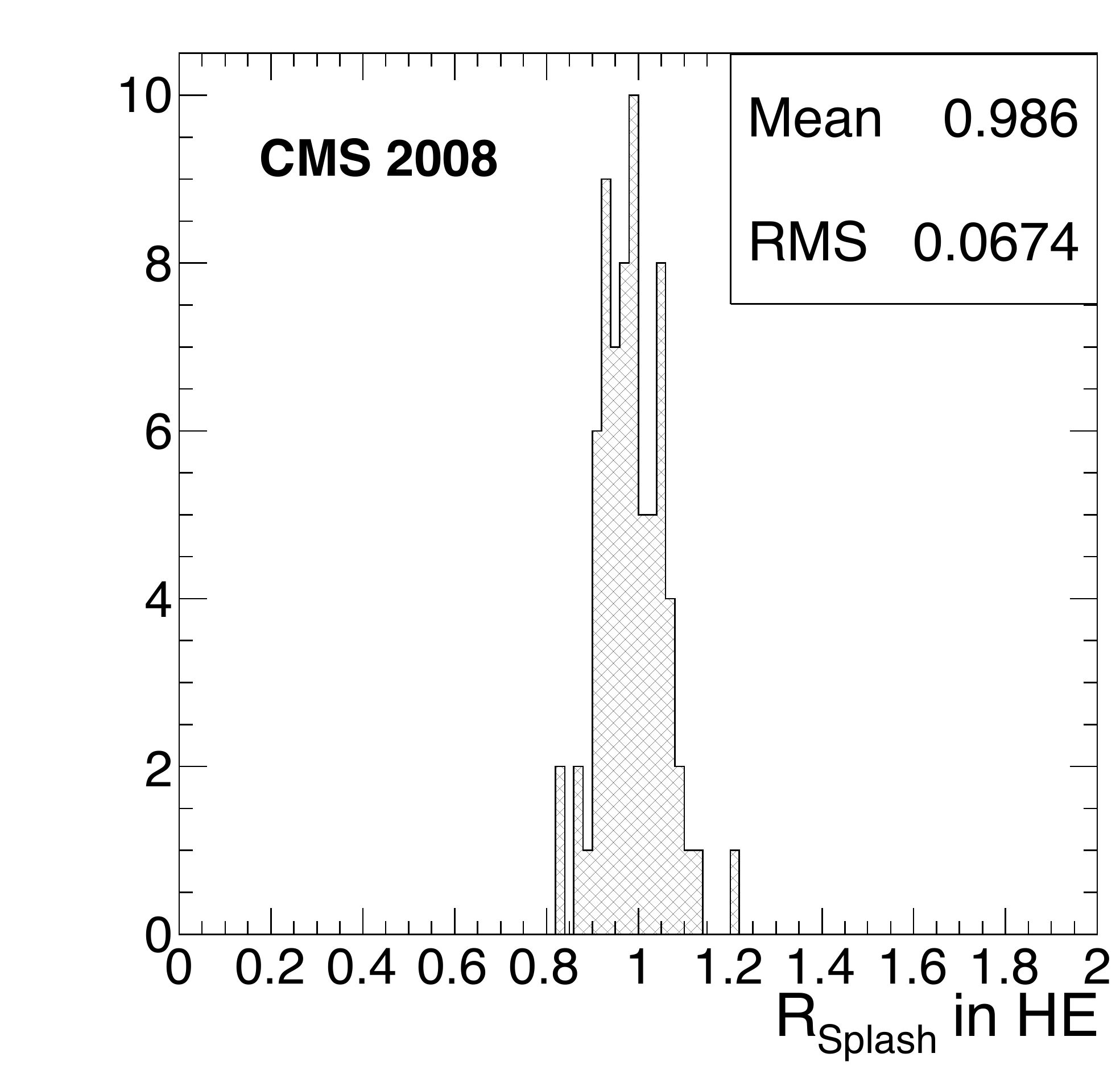}
    \hspace{1cm}
    \caption{Left: Ratio, of the splash muons, of the average signal
amplitudes  in the two HCAL endcaps, $-z$ and $+z$,  as a function
of the $\phi$ sector index, $i\phi$.  Right: Corresponding distribution of the signal ratios.
}
    \label{fig:he_validation}
  \end{center}
\end{figure}

Similar studies were performed for HE with a limited number of cosmic ray  muon events and all splash events.
The ratio of the muon signals for the two endcaps as a function of the $\phi$ sector index
is shown in Fig.~\ref{fig:he_validation}\,(right) for splash events.
The $R_{Splash}$  mean and the spread of this ratio
are shown in Fig.~\ref{fig:he_validation}\,(left).  The average is consistent with unity and
the spread is smaller than 7\%. Therefore,  no correction to the HE calibration was  made.

\begin{figure}[hbtp]
  \begin{center}
    \includegraphics[height=0.2575\textheight,width=0.45\textwidth]
{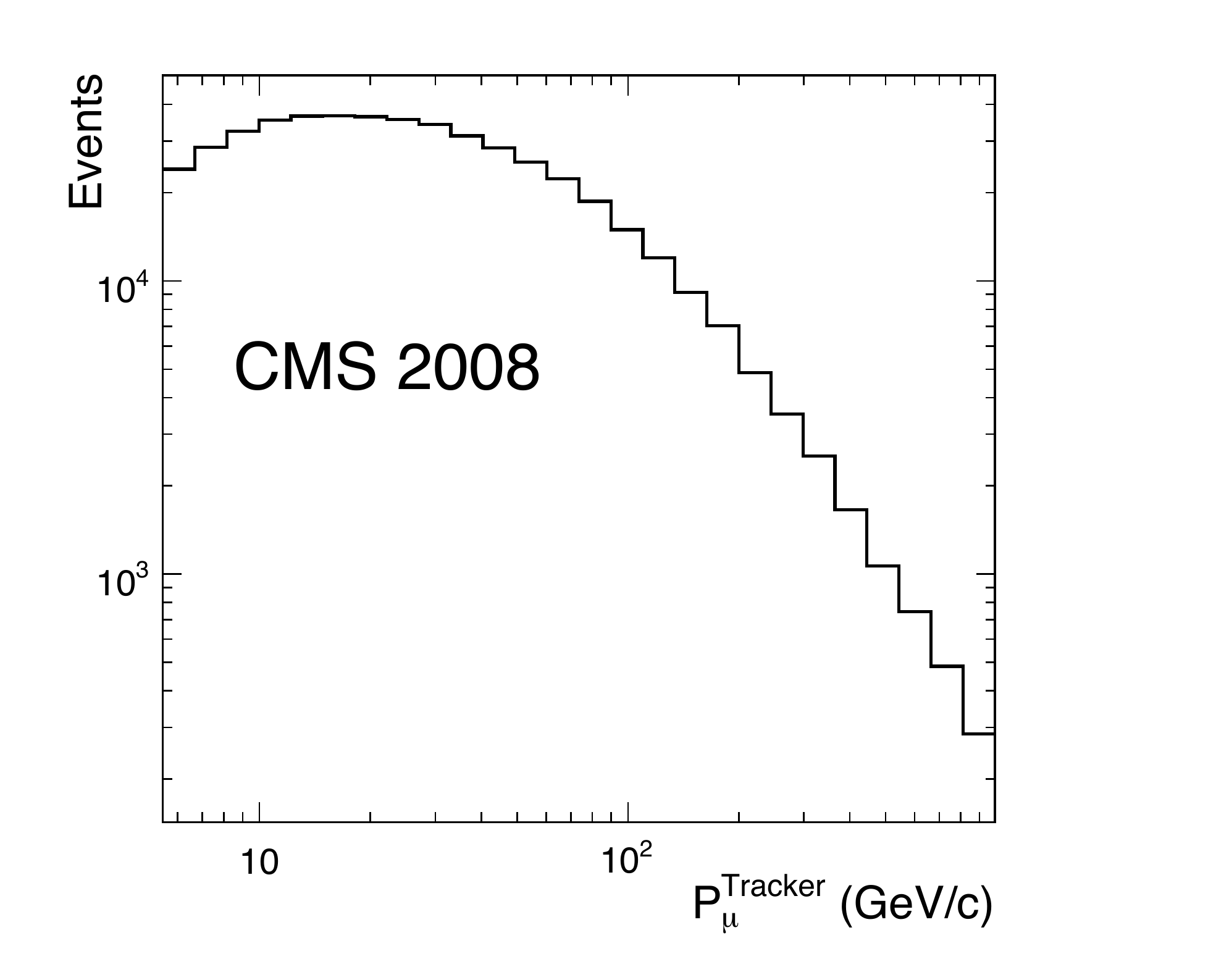}
\hspace{0.25cm}
    \includegraphics[height=0.265\textheight,width=0.52\textwidth]
{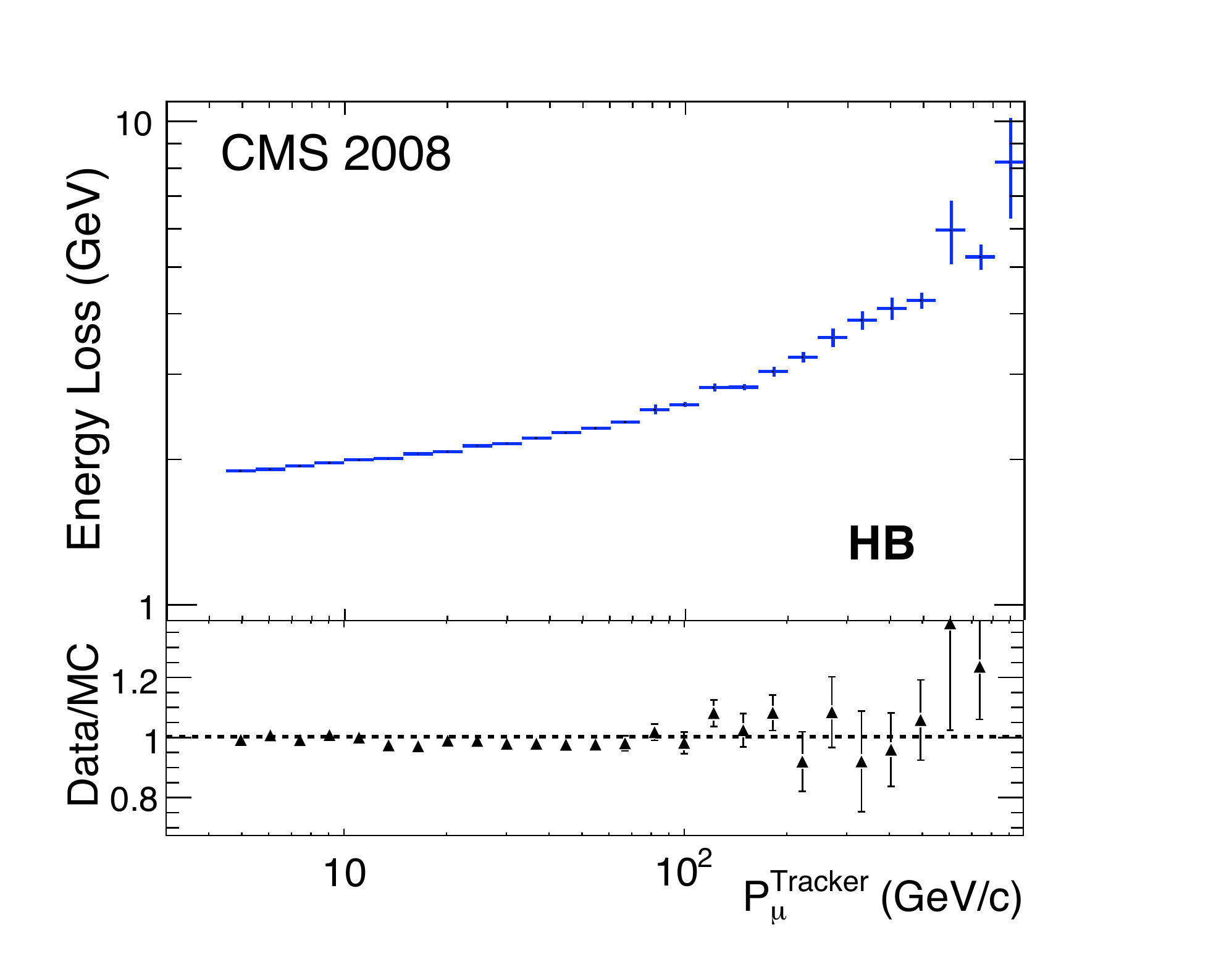}
    \hspace{1cm}
    \caption{ Left: Momentum spectrum of the muons selected by this  analysis.
Right:\,(top) Energy loss of cosmic  ray muons in HB after correcting the signal mean value for the muon path length
and normalizing to the thickness of HB at $\eta=0$; (bottom) ratio of data and  Monte Carlo simulation predictions\,(arbitrary scale)
for the bottom half of the calorimeter.
}
    \label{fig:dedx}
  \end{center}
\end{figure}

Cosmic ray muons were also used to validate the absolute energy scale and the material  description in
HB and HE simulations~\cite{bib:mc2}.  The absolute energy  scale was tested
 by comparing the energy loss of muons measured at a test beam and in CRAFT, at a fixed muon momentum.
From the test beam data it was derived that 150~GeV/$c$ muons deposited 2.80$\pm$0.03~GeV in
HB~\cite{bib:4}.
The energy loss measured from CRAFT data,  after taking
into account the magnetic field effects and the muon path length,
is 2.85~GeV with a statistical error of 0.02~GeV
for the momentum range between 135 and 170\,GeV/$c$.  This measurement is
in good agreement with the test beam result.
A wide muon momentum range was used to test the simulation.
Figure\,\ref{fig:dedx}\,(left) shows the cosmic ray  muon energy spectrum
measured with the tracker. Figure\,\ref{fig:dedx}\,(right top) shows the muon energy deposited in a calorimeter tower,
corrected for the path length and normalized to the thickness of HB at $\eta=0$, that is, 110\,cm.
The relativistic rise of the energy loss for muons between 6 and 1000\,GeV/$c$
is clearly observed  and is  consistent with expectations\,\cite{bib:pdg}.
Figure\,\ref{fig:dedx}\,(right bottom) illustrates the consistency
of the data and simulation~\cite{bib:mc2} as a function of the track momentum; the data to simulation
ratio shown at   an arbitrary scale.



\subsection{The Outer Hadron Calorimeter}

Since the HO scintillators
were not precalibrated with the radioactive source, the
calibration of each tower was performed using cosmic ray  muons
from CRUZET, at a uniform high voltage value of 8\,kV. This
calibration agrees well with the measurements performed at a test beam \cite{bib:4}.

\begin{figure}[hbtp]
  \begin{center}
    \includegraphics[width=0.4\textwidth]{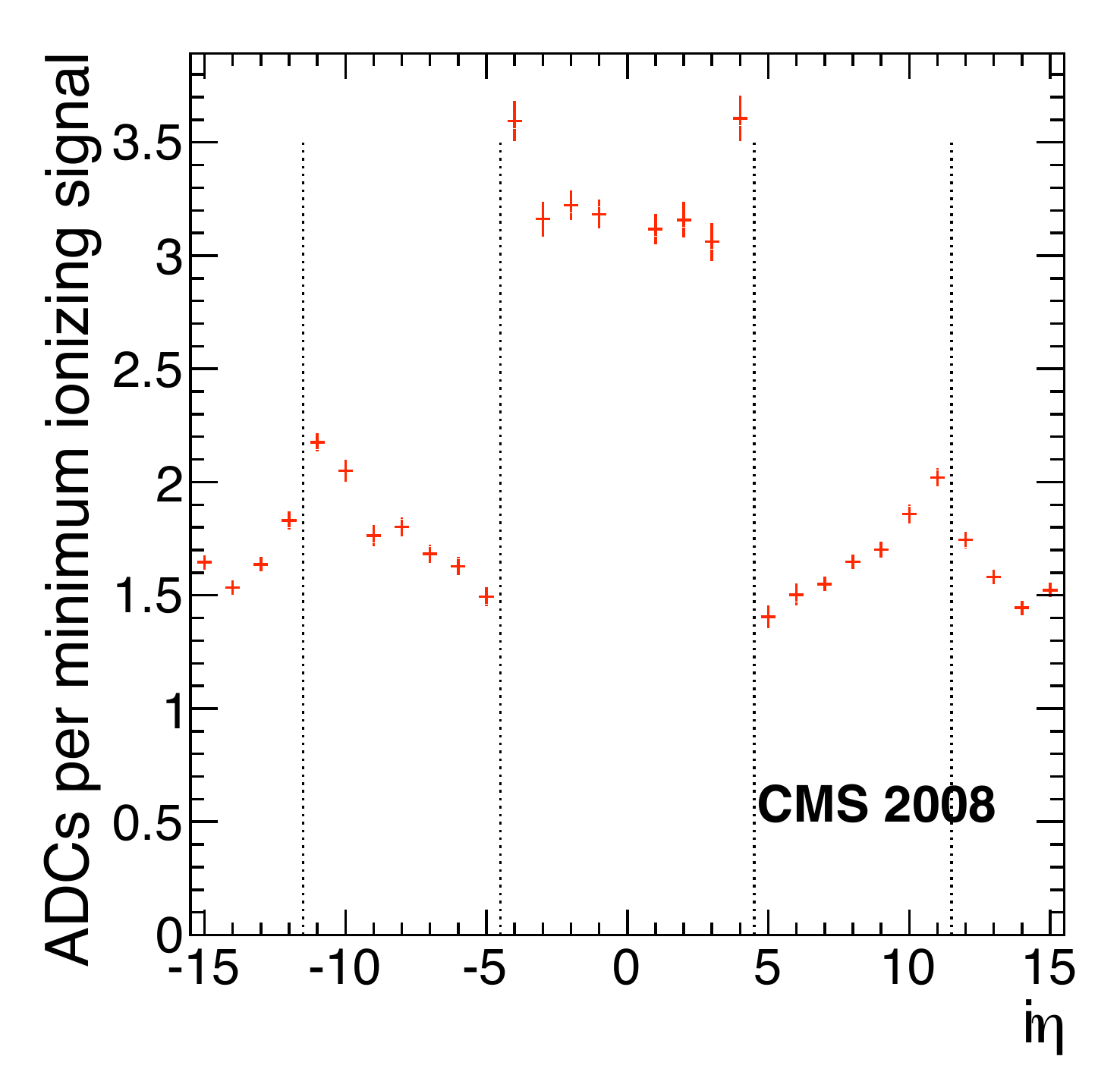}
    \hspace{1cm}
     \caption{Mean values of the signals
measured for a minimum ionizing muon   in HO as a function of the $\eta$ index, $i\eta$. The vertical lines indicate the borders of the rings.
The differences in signal size between Ring~0 and Rings~1 or  2 are due to
the fact that Ring~0 contains two layers of scintillators, for which
the signals are added optically.
The structure reflects the differences in the fiber lengths.}
    \label{fig:calibration_signals_hO}
  \end{center}
\end{figure}

Figure~\ref{fig:calibration_signals_hO} shows the mean values of  the
muon signals in HO, given in ADC counts for CRUZET data.
The structure within each
ring reflects the differences in  signal attenuation caused by the
varying length of the  fibers
that carry the signals to the photodetectors.



\subsection{The Forward Hadron Calorimeter}

\begin{figure}[hbtp]
  \begin{center}
    \includegraphics[width=0.45\textwidth]{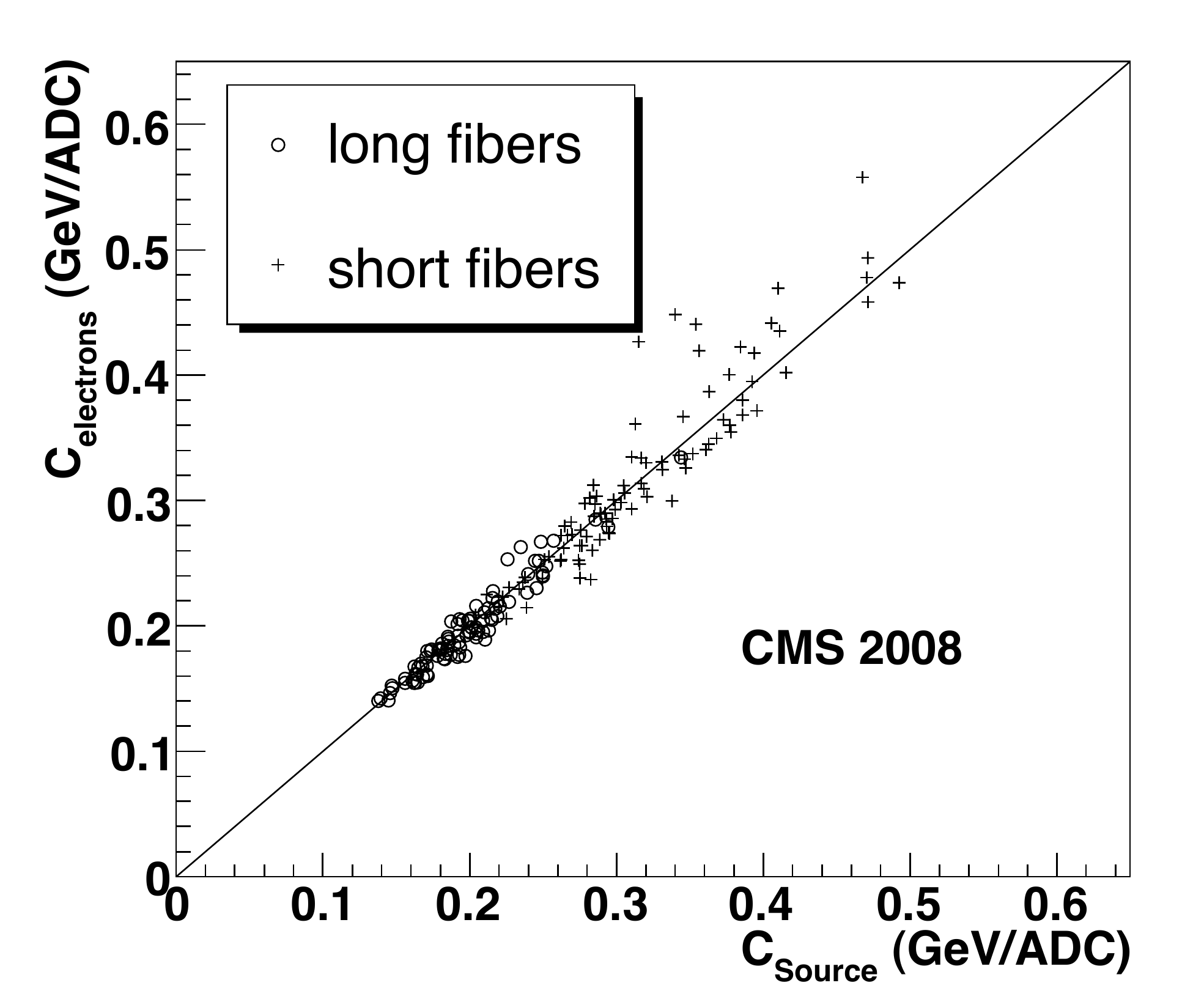}
    \hspace{1cm}
  \caption{The signal in GeV/ADC for 100 GeV/$c$ electrons  from the test beam\,(vertical axis)
versus the radioactive source in GeV/ADC for the same $\eta$, $\phi$ space, for short\,(crosses)
and long\,(circles) fibers for HF$+$ and HF$-$.  This information was used
to set the energy scale of the calibration obtained with the radioactive source method.
}
  \label{fig:hfsource}
  \end{center}
\end{figure}

The calibration for HF was obtained from measurements with radioactive
sources and from test beam data.
Five HF wedges were precalibrated at the test beam with 100\,GeV/$c$ electrons~\cite{bib:1}
and with a
5\,mCi Co$^{60}$ source.  Figure~\ref{fig:hfsource} displays the good
correlation between the signals of the electrons
and the source measurement in the same $\eta$, $\phi$ sector, separately for
short\,(crosses) and long\,(circles) fibers.
The  average value for the calibration coefficients of the  short\,(long)
fibers is 0.34\,(0.21) GeV/fC.
This information was used to set the energy scale for
all the channels which were not exposed to the beam, using the results
of their measurement with the radioactive source.

The monitoring of the HF calibration relies on single photoelectron
distributions\,(SPE) measuments performed with the LED system.  Earlier measurements performed at a   test beam
showed that 25 SPEs are equivalent to 100 GeV~\cite{bib:1}. The major difference
between using the SPE measurements and using
a radioactive source is that the SPE characterizes the PMT performance
alone, while  the radioactive source creates Cherenkov light inside the detector which in turn detected by the PMTs.
Therefore, the light collection efficiency and the PMT photocathode sensitivity are taken into account in the source calibration,
but not in the SPE technique.  However, if the attenuation of Cherenkov light as it propagates from
the fiber to the PMT is only affected by small variations from channel to channel,
 or if the attenuation is negligible,
the two calibration techniques should be equivalent.

\begin{figure}[hbtp]
  \begin{center}
\includegraphics[width=0.415\textwidth]{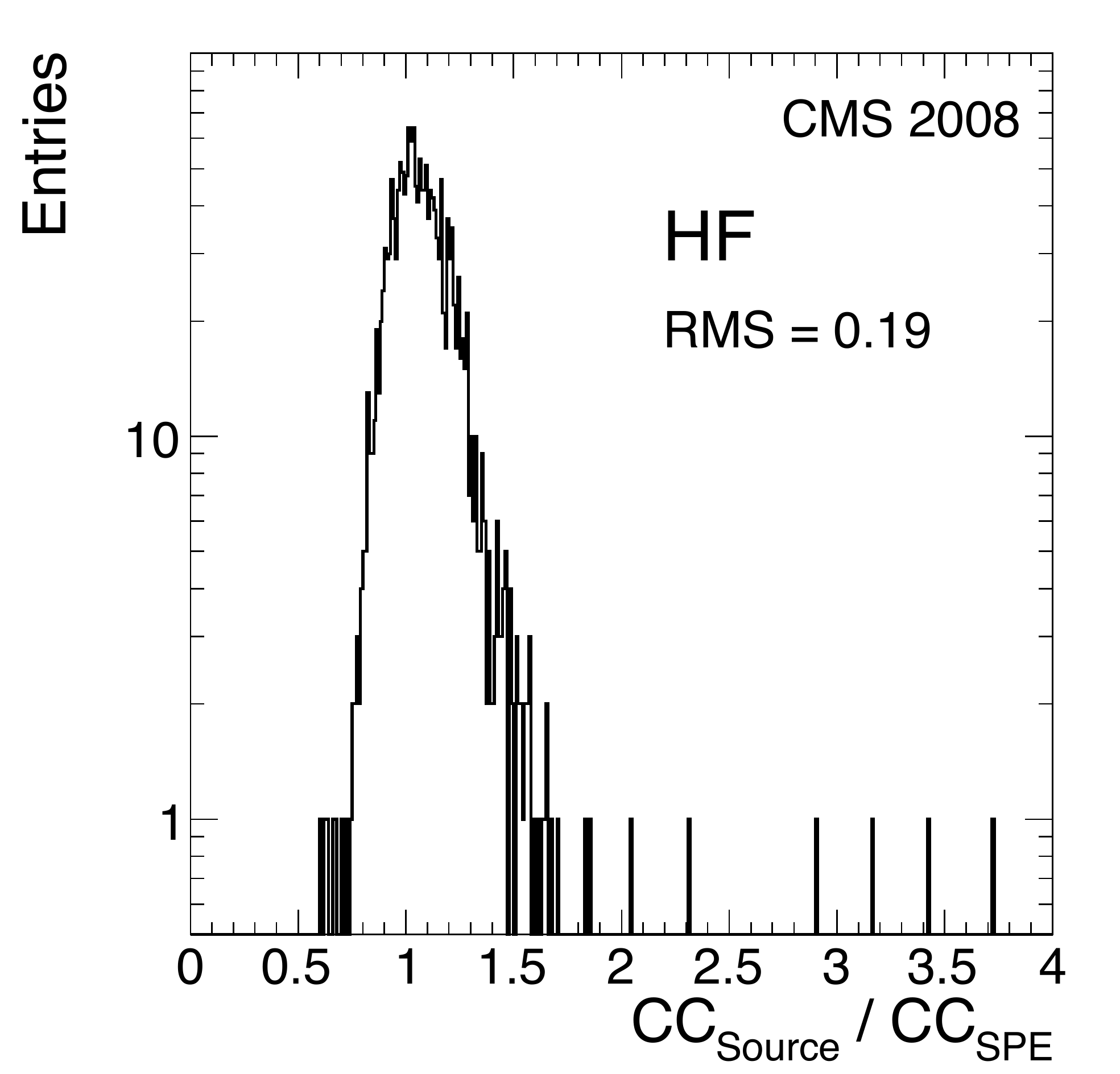}
\hspace{1cm}\includegraphics[width=0.415\textwidth]{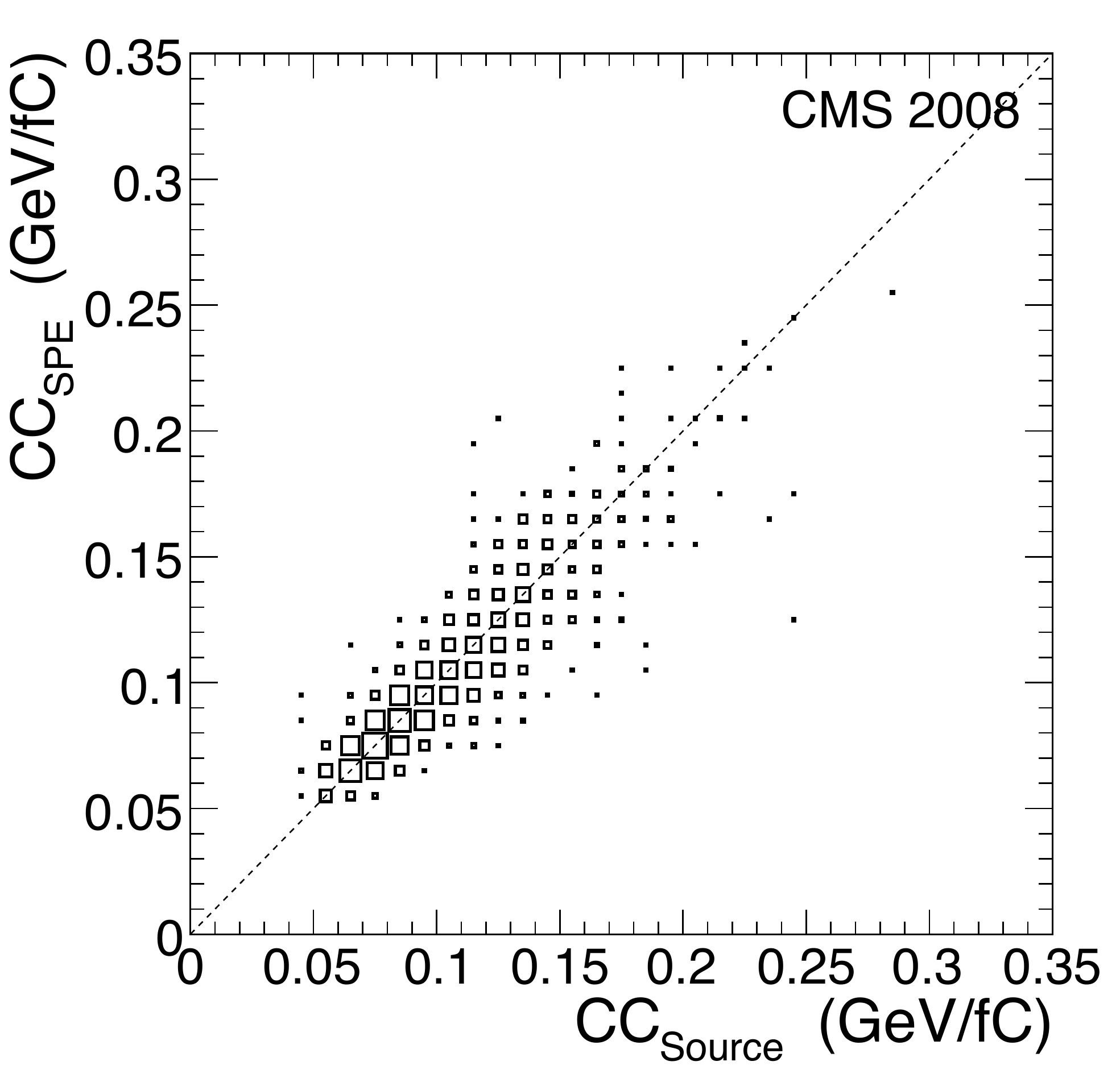}
    \hspace{1cm}
  \caption{The calibration coefficients\,(CC) for HF$+$ and HF$-$ obtained
    with the SPE\,(single photoelectron) method are compared to the standard values obtained with
the radioactive source method before\,(left plot shows ratio) and after\,(right plot  shows correlation)
correcting the calibration, see text.
The  ratio of the two sets of constants is around 1.1 because the measurements were performed
with a different HV setup.
}
  \label{fig:spevssource}
  \end{center}
\end{figure}

A comparison between the results obtained with the radioactive source and new
SPE measurements, performed after the final HF installation,
revealed that 20 of the 1728  channels had inconsistent calibration,
see  Fig.~\ref{fig:spevssource}~(left).  The beam splash events
confirmed this calibration problem since the same channels were found to
show an energy response inconsistent with their neighbors.
Therefore, for those channels the source calibration constants were replaced
by    the SPE  constants.

Figure~\ref{fig:spevssource}\,(right) shows the correlation between the calibration coefficients obtained
from single photoelectron distributions and the coefficients obtained from the radioactive source calibration,
after correcting the few outliers mentioned above.

\section{Summary}

The performance of the CMS hadron calorimeter
has been studied using cosmic ray and beam splash
events, providing improved calibrations with
respect to those obtained in test beam data and
with radioactive sources.

As a result of this calibration effort, the barrel
portion is intercalibrated to the level of 5\%,
for 85\% of the channels.  The other HCAL portions
have all their channels intercalibrated to better
than 10\% (HE), 12\% (HF) and 5\% (HO).
Furthermore, the presence
of the magnetic field results in a measured increase of the
signal, by 9.4\% in HB and 8\% in HE, as expected.

Noise studies have been performed, showing
that noise will contribute a trigger rate of
around 100 Hz. This rate is stable.

In conclusion, the CRAFT data and the
beam splash events from 2008 provided
an improved calibration of the HCAL.
This subdetector is now ready
for LHC collisions.

\section*{Acknowledgments}

We thank the technical and administrative staff at CERN and other CMS Institutes,
and acknowledge support from:
FMSR (Austria);
FNRS and FWO (Belgium);
CNPq, CAPES, FAPERJ, and FAPESP (Brazil);
MES (Bulgaria);
CERN;
CAS, MoST, and NSFC (China);
COLCIENCIAS (Colombia);
MSES (Croatia);
RPF (Cyprus);
Academy of Sciences and NICPB (Estonia);
Academy of Finland, ME, and HIP (Finland);
CEA and CNRS/IN2P3 (France);
BMBF, DFG, and HGF (Germany);
GSRT (Greece);
OTKA and NKTH (Hungary);
DAE and DST (India);
IPM (Iran);
SFI (Ireland);
INFN (Italy);
NRF (Korea);
LAS (Lithuania);
CINVESTAV, CONACYT, SEP, and UASLP-FAI (Mexico);
PAEC (Pakistan);
SCSR (Poland);
FCT (Portugal);
JINR (Armenia, Belarus, Georgia, Ukraine, Uzbekistan);
MST and MAE (Russia);
MSTDS (Serbia);
MICINN and CPAN (Spain);
Swiss Funding Agencies (Switzerland);
NSC (Taipei);
TUBITAK and TAEK (Turkey);
STFC (United Kingdom);
DOE and NSF (USA).
Individuals have received support from
the Marie-Curie IEF program (European Union);
the Leventis Foundation;
the A. P. Sloan Foundation;
and the Alexander von Humboldt Foundation.

\bibliography{auto_generated}   
\clearpage

\cleardoublepage\appendix\section{The CMS Collaboration \label{app:collab}}\begin{sloppypar}\hyphenpenalty=500\textbf{Yerevan Physics Institute,  Yerevan,  Armenia}\\*[0pt]
S.~Chatrchyan, V.~Khachatryan, A.M.~Sirunyan
\vskip\cmsinstskip
\textbf{Institut f\"{u}r Hochenergiephysik der OeAW,  Wien,  Austria}\\*[0pt]
W.~Adam, B.~Arnold, H.~Bergauer, T.~Bergauer, M.~Dragicevic, M.~Eichberger, J.~Er\"{o}
% ...
\vskip\cmsinstskip
% ...

\end{sloppypar}
\end{document}